\documentclass[aps,twocolumn,titlepage,nofootinbib]{revtex4}
\usepackage{graphicx,physics}

\usepackage[utf8]{inputenc}

\usepackage[normalem]{ulem}
\usepackage{hyperref,amssymb}
\usepackage{url}
\usepackage{color,soul}
\usepackage{graphicx}
\usepackage{verbatim}
\usepackage{multirow}
\usepackage{amsmath}
\newcommand{\beq}{\begin{eqnarray}}
\newcommand{\eeq}{\end{eqnarray}}
\usepackage{mathrsfs}
\usepackage{float,soul}
\usepackage[dvipsnames]{xcolor}
\usepackage{mathtools}
\usepackage{slashed}
\usepackage{physics}	
\usepackage{graphicx}   
\usepackage{epstopdf}
\usepackage{subfigure}  
\usepackage{bbold}
\usepackage{wasysym}
\usepackage{feynmp}
\usepackage{hyperref}
\usepackage{tabularx}

\hypersetup{colorlinks,
}
\usepackage{bm}

\begin{document}

\title{ \large A flat-band perspective on the boson peak in amorphous solids}
\author{Shivam Mahajan$^1$}
\author{Long-Zhou Huang$^{2,3}$}
\author{Cunyuan Jiang$^{4,5}$}
\author{Yun-Jiang Wang$^{2,3}$}
\email{yjwang@imech.ac.cn}
\author{Massimo Pica Ciamarra$^1,^6$}
\email{massimo@ntu.edu.sg}
\author{Jie Zhang$^{4,7}$}
\email{jiezhang2012@sjtu.edu.cn }
\author{Matteo Baggioli$^{4,5}$}
\email{b.matteo@sjtu.edu.cn}
\address{$^1$Division of Physics and Applied Physics, School of Physical and
Mathematical Sciences, Nanyang Technological University, Singapore}
\address{$^2$State Key Laboratory of Nonlinear Mechanics, Institute of Mechanics, Chinese Academy of Sciences, Beijing 100190, China}
\address{$^3$School of Engineering Science, University of Chinese Academy of Sciences, Beijing 100049, China}
\address{$^4$School of Physics and Astronomy, Shanghai Jiao Tong University, Shanghai 200240, China}
\address{$^5$Wilczek Quantum Center, School of Physics and Astronomy, Shanghai Jiao Tong University, Shanghai 200240, China}
\address{$^6$CNR--SPIN, Dipartimento di Scienze Fisiche, Universit\`a di Napoli Federico II, I-80126, Napoli, Italy} 
\address{$^7$Institute of Natural Sciences, Shanghai Jiao Tong University, Shanghai 200240, China}

\begin{abstract}
The boson peak is a characteristic anomaly of amorphous solids broadly defined as a low-energy excess in the density of states and heat capacity compared to the textbook predictions of Debye theory. The origin of this anomaly has long been the subject of ongoing debate and remains a topic of active controversy. We propose that the boson peak may have a defining dynamical feature: the accumulation of vibrational spectral weight within a narrow frequency window that is only weakly dependent on wavevector. In this perspective, the boson peak reflects a flat or weakly dispersive band in the dynamical structure factor rather than a propagating excitation.
We revisit both experimental and simulation data from the literature through this lens and conduct further simulations in 2D and 3D amorphous systems. Taken together, these analyses provide compelling converging evidence for this interpretation and sharply constrain the space of viable theoretical descriptions of the boson peak.
\end{abstract}


\maketitle

\section*{Introduction}
\begin{figure}[!t]
\centering
    \includegraphics[width=0.65\linewidth]{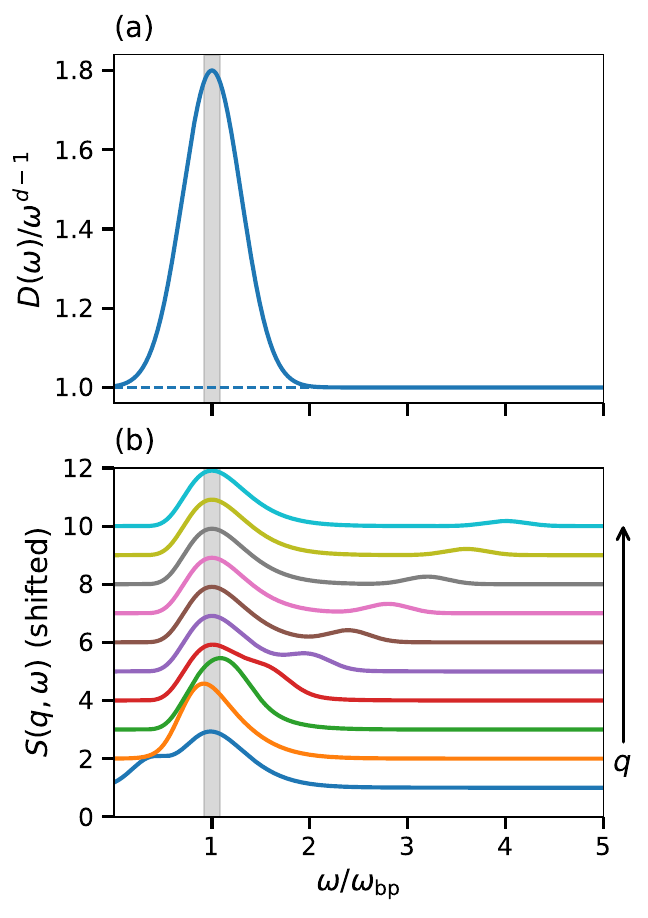}
    \caption{
   \textbf{(a)} Schematic representation of the boson peak as conventionally identified from a peak in the reduced vibrational density of states.
    \textbf{(b)} Schematic dynamic structure factor illustrating the viewpoint that the boson peak is associated with a flat band in $S(q,\omega)$, a perspective that provides constraints on theoretical descriptions of its physical origin.
    }
    \label{fig:0}
\end{figure}

Our understanding of the vibrational density of states (VDOS) and heat capacity of ideal crystalline materials is rooted in a century-old theory: the Debye model \cite{https://doi.org/10.1002/andp.19123441404}. 
This well-established paradigm is based on the premise that the only relevant low-energy excitations in crystals are propagating collective plane waves, also named acoustic phonons. 
Neglecting dissipation, which would otherwise lead to attenuation, and ignoring different polarizations and possible anisotropies, these sound waves follow a linear dispersion relation $\omega = v q$ (see black dashed line in Fig. \ref{fig:0}), where $\omega$, $q$, and $v$ denote the frequency, wavevector, and propagation speed, respectively.

The properties of these acoustic phonons can be derived either through a microscopic normal mode analysis \cite{moon2023normalmodedecompositionatomic}, or more abstractly, from the spontaneous breaking of translational symmetry \cite{Leutwyler:1996er}, that is, the formation of a periodic, ordered lattice. 
Leaveraging on these concepts, the vibrational density of states of ideal crystals takes the well-known form: 
\begin{equation}
    g(\omega) \propto \omega^{D-1}, \label{debyelaw}
\end{equation}
where $D$ is the number of spatial dimensions and numerical prefactors are omitted for simplicity. Equation~\ref{debyelaw}, referred to hereafter as the Debye law, holds in the low-frequency regime where the phonon dispersion remains linear. In the Debye approximation, this linearity is assumed to extend up to a maximum cutoff frequency, the Debye frequency $\omega_D$, beyond which $g(\omega)$ vanishes.

By further recognizing that phonons are quantum mechanical in nature, specifically, bosonic excitations, Debye also derived an expression for the heat capacity of crystals, which in the low-temperature limit takes the universal form:
\begin{equation}
    C(T) \propto T^D. \label{debyelaw2}
\end{equation}
Both of Debye's predictions have been experimentally confirmed in crystalline materials and stand as one of the most successful paradigms in solid-state physics \cite{kittel2018introduction}.

However, in 1971, Zeller and Pohl \cite{PhysRevB.4.2029} reported that noncrystalline solids present characteristic anomalies in their heat capacity that evade the Debye paradigm. More precisely, they observed an anomalous peak in the normalized heat capacity $C(T)/T^3$ around $\approx 10 $ K. 
This feature is now known as the ``\textit{boson peak}'' (BP). 
Moreover, they speculated that this departure from the Debye specific heat may be characteristic of the glassy state, or of amorphous solids in general. Later, in the mid-1980s, Buchenau and colleagues \cite{PhysRevLett.53.2316,PhysRevB.34.5665} convincingly demonstrated that this anomaly was not related to the Bose-Einstein distribution, from where it takes its name, but rather stemmed from an excess of vibrational modes visible in the density of states, exceeding the prediction of the Debye law, as schematically illustrated in Fig.~\ref{fig:0}(a). 
This finding was subsequently confirmed by inelastic neutron scattering and various other experimental techniques (e.g., \cite{PhysRevLett.84.5355,doi:10.1126/science.280.5369.1550,PhysRevLett.92.245508,PhysRevB.94.224204}), and it is now widely accepted as an established fact.

Despite more than half a century of experimental investigations and theoretical modeling, the origin of the boson peak in amorphous solids remains highly controversial and often the subject of debate. We refer to the excellent book by Ramos \cite{ramos2022low} for an extensive review of the \textit{status quo}.

This persistent debate highlights the need for new, physically grounded ways of constraining theoretical descriptions, in order to assess whether a single framework can account for the boson peak across different amorphous systems, or whether distinct mechanisms are required to describe different classes of materials.

A possible constraint is suggested by recent observations that the boson peak is associated with a nearly dispersionless feature in the dynamical structure factor $S(q,\omega)$ \cite{tanakaNatPhys}: in simulated glassy systems, the dynamical structure factor exhibits a band in the $(q,\omega)$ plane where spectral weight accumulates around a characteristic frequency $\omega \simeq \omega_0$, with only a weak dependence on the wavevector $q$, as schematically illustrated in Fig.~\ref{fig:0}(b).
The key result of Refs.~\cite{tanakaNatPhys,tanakaPRR} is that this characteristic frequency $\omega_0$ coincides with the boson-peak frequency $\omega_{\mathrm{bp}}$ in both two- and three-dimensional systems.

We highlight that a weakly dispersive concentration of spectral weight naturally produces a pronounced peak in the vibrational density of states, potentially broadened by disorder or thermal effects, independently of the microscopic origin.
Indeed, within the harmonic approximation, the vibrational density of states follows, schematically, $g(\omega)\propto \omega^2 \int d^Dq\, S(q,\omega)/q^2$. A flat or weakly dispersive band in the $(q,\omega)$ plane therefore implies that a large region of reciprocal space contributes intensity at essentially the same frequency. Integrating over this region necessarily yields an excess in $g(\omega)$ at around the band frequency. 
This mechanism is directly analogous to the formation of van Hove singularities in crystalline solids, although here it applies to a broadened, disorder-smeared concentration of spectral weight rather than to well-defined phonon branches.

The presence of a flat or weakly dispersive band in $S(q,\omega)$ has important physical implications. It indicates a collective dynamical process that spans a broad range of length scales while remaining confined to a narrow frequency window, and it imposes a stringent constraint on theoretical descriptions, which must account for both the absence of a characteristic propagation length and the robustness of the associated frequency scale. If such a non-phononic flat band at the boson-peak frequency were indeed a robust feature of amorphous solids, the central question would shift to understanding the microscopic origin of this dispersionless spectral structure. It is therefore essential to establish whether this phenomenology is universal and whether it has already been observed, explicitly or implicitly, in other systems, including experimental studies.

Interestingly, as noted in the \textit{Introduction} of \cite{ramos2022low}, as early as 1959, prior to the seminal work of Zeller and Pohl, Flubacher et al. \cite{FLUBACHER195953} suggested that the excess heat capacity in vitreous silica was due to low-frequency optical-like (i.e., dispersionless) modes detected via Raman scattering. This idea was later revisited in depth by Nakayama; see \cite{Tsuneyoshi_Nakayama_2002} for a comprehensive review.

In this work, we investigate whether the boson peak is universally associated with the emergence of a flat-band.
We support this scenario with substantial evidence, combining new numerical analyses with a re-examination of existing experimental data and prior results on the dynamical structure factor.

Specifically, (i) we show that this phenomenon was present, but previously overlooked, in experimental data on a two-dimensional (2D) jammed packing of photoelastic disks, (ii) we show that this feature was already evident in the experimental dynamical structure factor of several disordered systems ranging from granular packings and silicate glasses, to metallic glasses and polymers, (iii) we provide additional simulation evidence for this scenario in systems of polydisperse particles interacting via a Lennard-Jones-like potential in both 2D and 3D, along with a simulated 3D Cu-Zr metallic glass, and (iv) we confirm a recently discussed correlation between the intensity of this flat band and the static structure factor \cite{Li2025}.
Finally, we critically examine how the existence of a flat band constrains several popular theoretical scenarios proposed for the boson peak.

\section*{Methods and Terminology}
The central quantity of our analysis is the dynamic structure factor. Specifically, we consider the proper longitudinal $S_L(q,\omega)$ dynamic structure factor, which describes how density fluctuations evolve in space and time, and its transverse generalization, $S_T(q,\omega)$.

In numerical simulations, these quantities can be accessed by first performing an eigenmode decomposition and then evaluating, for each eigenmode $\mathbf{e}_\lambda$ with frequency $\omega_\lambda$, the polarization-resolved weights
\begin{align*}
&E_{\lambda,\text{T}}(\mathbf{q}) = \left| \sum_j [ \hat{\mathbf{q}} \times \mathbf{e}_{\lambda}(j) ] \exp\!\left( i\,\mathbf{q}\cdot\mathbf{r}_j \right) \right|^2,\\
&E_{\lambda,\text{L}}(\mathbf{q}) = \left| \sum_j [ \hat{\mathbf{q}} \cdot \mathbf{e}_{\lambda}(j) ] \exp\!\left( i\,\mathbf{q}\cdot\mathbf{r}_j \right) \right|^2.
\end{align*}
Here $\hat{\mathbf{q}}$ is the unit wave vector, $M$ is the effective particle mass, and $\mathbf{r}_j$ is the equilibrium position of particle $j$. One then evaluates
\begin{equation}
S_{\alpha}(q,\omega) = \frac{k_B T}{M}\,\frac{q^2}{\omega^2} \sum_{\lambda} E_{\lambda,\alpha}(\mathbf{q})\,\delta(\omega-\omega_\lambda),
\end{equation}
where $\alpha=T,L$.

In colloidal and granular scale experiments, both $S_L$ and $S_T$ are experimentally accessible, as particle motion can be resolved in time. In contrast, in molecular systems, the scattering intensity
\begin{equation}
I(q,E)\propto S_L(q,E)\,e^{-2W(q)}, \qquad (E=\hbar\omega),
\end{equation}
measures the longitudinal dynamic structure factor, attenuated by the Debye--Waller factor $e^{-2W(q)}$.

At low frequencies, the polarization-resolved spectra $S_\alpha(q,\omega)$ contain the characteristic $\omega^{-2}$ acoustic background arising from longitudinal and transverse phonons. To remove this background and expose the excess modes underlying the boson peak, we consider a rescaled inelastic-scattering spectrum\footnote{In the experimental context~\cite{Li2025}, the definition of $B(q,E=\hbar\omega)$ includes Bose, Debye–Waller, and kinematic factors that must be removed from measured intensities; in our theoretical setting, these prefactors are absent, and the simpler form of \eqref{eq:reducedSq} suffices.}
\begin{equation}
B_\alpha(q,\omega)=\omega^{2}\,S_\alpha(q,\omega).
\label{eq:reducedSq}
\end{equation}
At fixed $\omega$, $B_\alpha(q,\omega)$ then reveals the momentum-space structure of the non-phononic vibrational modes contributing at that frequency.

Due to the widespread expectation that the boson peak is predominantly associated with transverse dynamics, most of the investigations we review---as well as our own analyses---focus on the transverse dynamical structure factor $S_T(q,\omega)$. Neutron scattering measurements, however, provide access only to the longitudinal component $S_L(q,\omega)$, and thus represent a complementary probe. For completeness, we report $S_L(q,\omega)$ for the model systems we have studied in the Supporting Information (SI) and verify that, in these cases, the boson--peak anomaly indeed appears primarily in the transverse channel.

\section*{Review of existing simulation evidence}
\begin{figure}[!t]
    \includegraphics[width=\linewidth]{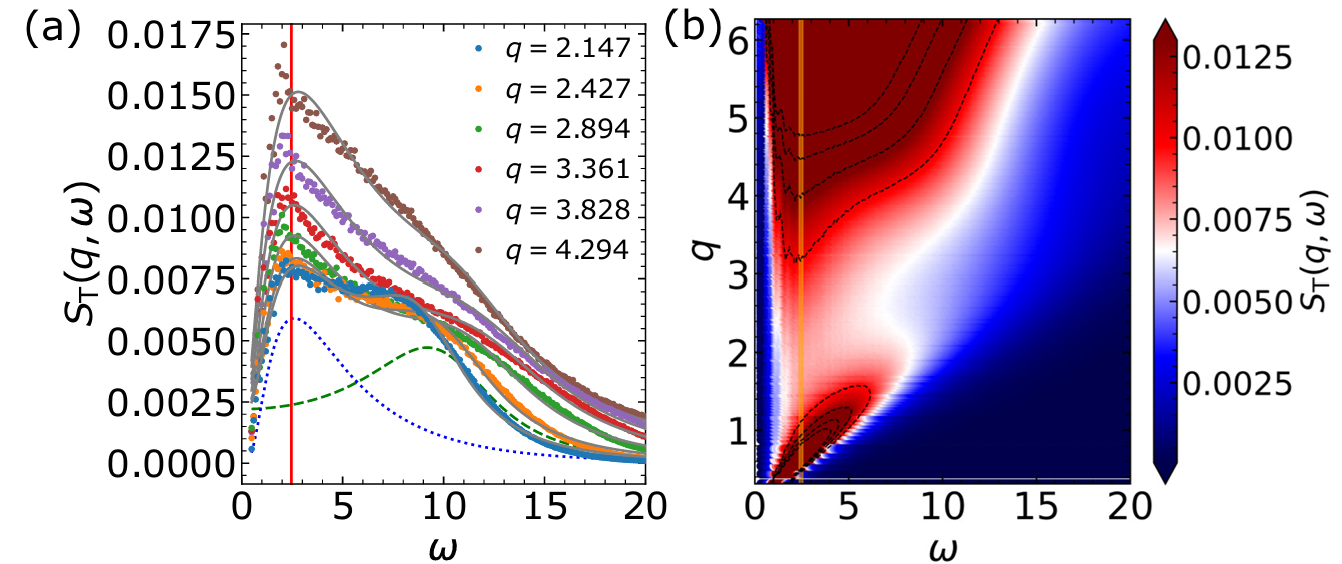}
    \caption{Transverse dynamic structure factor $S_T(q,\omega)$ of simulated 3DIPL glasses. \textbf{(a)} 1D representation along constant $q$ cuts; \textbf{(b)} 2D color map. Vertical red and orange lines indicate the BP frequency as obtained from the reduced density of states. Blue dotted and green dashed lines correspond respectively to the contribution of the flat dispersionless mode and the transverse acoustic phonons for $q=2.427$. Figure adapted with permission from Ref.~\cite{tanakaPRR}.}
    \label{fig:tanaka}
\end{figure}
Despite similar ideas having surfaced in the past, the most compelling simulation evidence for the connection between the boson peak and a dispersionless flat band has emerged only recently, in a series of works by Hu and Tanaka~\cite{tanakaNatPhys, tanakaPRR}, which have strongly inspired the explorations presented in this Perspective. Following the seminal work by Shintani and Tanaka \cite{Shintani2008}, these studies reframe the boson peak not as a mere excess in the vibrational density of states $g(\omega)$, but as a spatially resolved feature in the dynamic structure factor $S(q,\omega)$, as illustrated in Fig.~\ref{fig:0}.

In~\cite{tanakaNatPhys}, extensive molecular dynamics simulations were carried out for a two-dimensional bidisperse system in which particles interact via a power-law potential, $V(r) \propto r^{-10}$. This analysis was later extended in~\cite{tanakaPRR} to three-dimensional (3D) model glasses with isotropic repulsive and attractive interactions.

A key observation from both studies is that the dynamic structure factor for transverse fluctuations, $S_T(q,\omega)$, with $q$ and $\omega$ denoting the wavevector and frequency, respectively, cannot be described solely by acoustic phonons modeled via a damped harmonic oscillator (see Fig.~\ref{fig:tanaka}(a)). Instead, a distinct contribution clearly emerges above a certain wavevector. Remarkably, this contribution was captured by an asymmetric peak structure, well-fitted using a log-normal function without any evident wavevector dependence. By incorporating both phononic and non-phononic contributions, the numerical dynamic structure factor could be accurately reconstructed.

The central result of this analysis is that the energy associated with this excess signal is independent of $q$ and aligns almost exactly with the boson peak frequency, as independently extracted from the Debye-reduced density of states (see Fig.~\ref{fig:tanaka}(b)). 
Setting aside the microscopic interpretation proposed in those studies, these empirical findings strongly suggest that the BP stems from a dispersionless non-phononic band. 

\begin{figure}[!t]
    \centering
    \includegraphics[width=\linewidth]{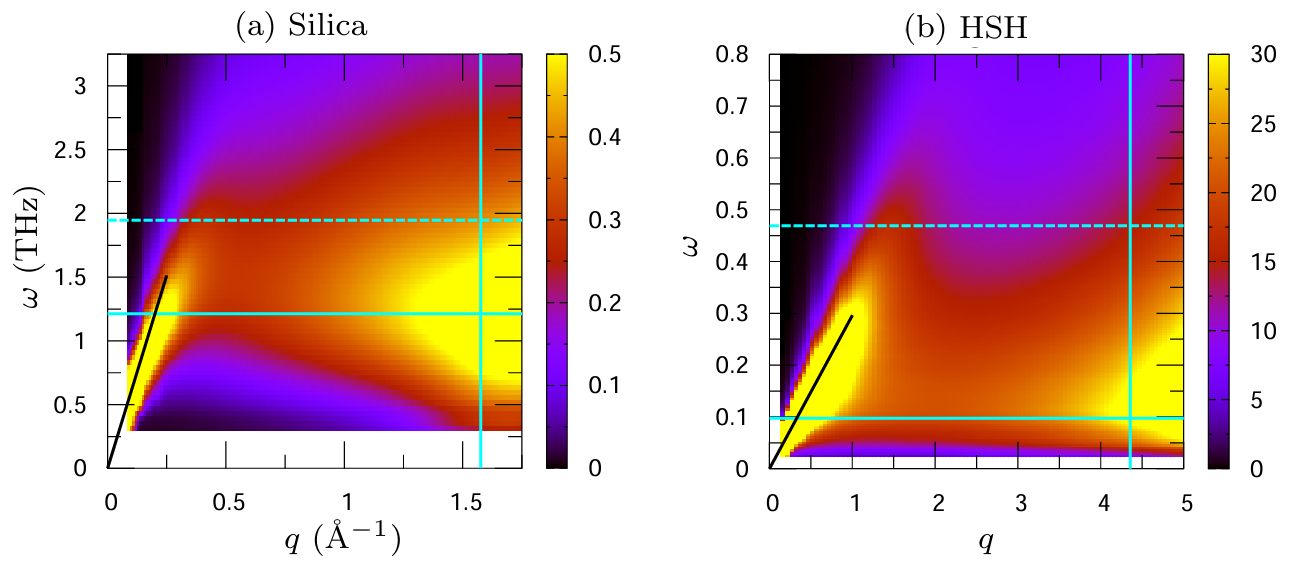}
    \caption{Transverse dynamic structure factor for silica glass \textbf{(a)} and a high-connectivity harmonic spheres (HSH) model \textbf{(b)}. The black solid curve is the linear dispersion of the acoustic phonon. The horizontal cyan solid line indicates the BP frequency. Figures adapted from Ref.~\cite{mizuno2025bosonpeakcovalentnetwork} with permission from the authors.}
    \label{fig:nn}
\end{figure}

Very recently, Ref.~\cite{mizuno2025bosonpeakcovalentnetwork,mizuno2026bosonpeakdynamicalstructure} performed extended numerical simulations for Silica (SiO$_2$) glass, harmonic sphere (HS) glasses and Lennard-Jones (LJ) glass. For all these systems, a wavenumber-independent non-phononic band was observed in the dynamical structure factor at the BP energy (see Fig.~\ref{fig:nn} for a sample of their results).

In the following, we present simulation and experimental evidence suggesting that the relation between BP and a dispersionless mode is universal, i.e., independent of specific interactions, compositions, or system details.

\section*{Novel evidence in 2D and 3D simulated glasses
}

\begin{figure*}[!t]
    \centering
    \includegraphics[width=\linewidth]{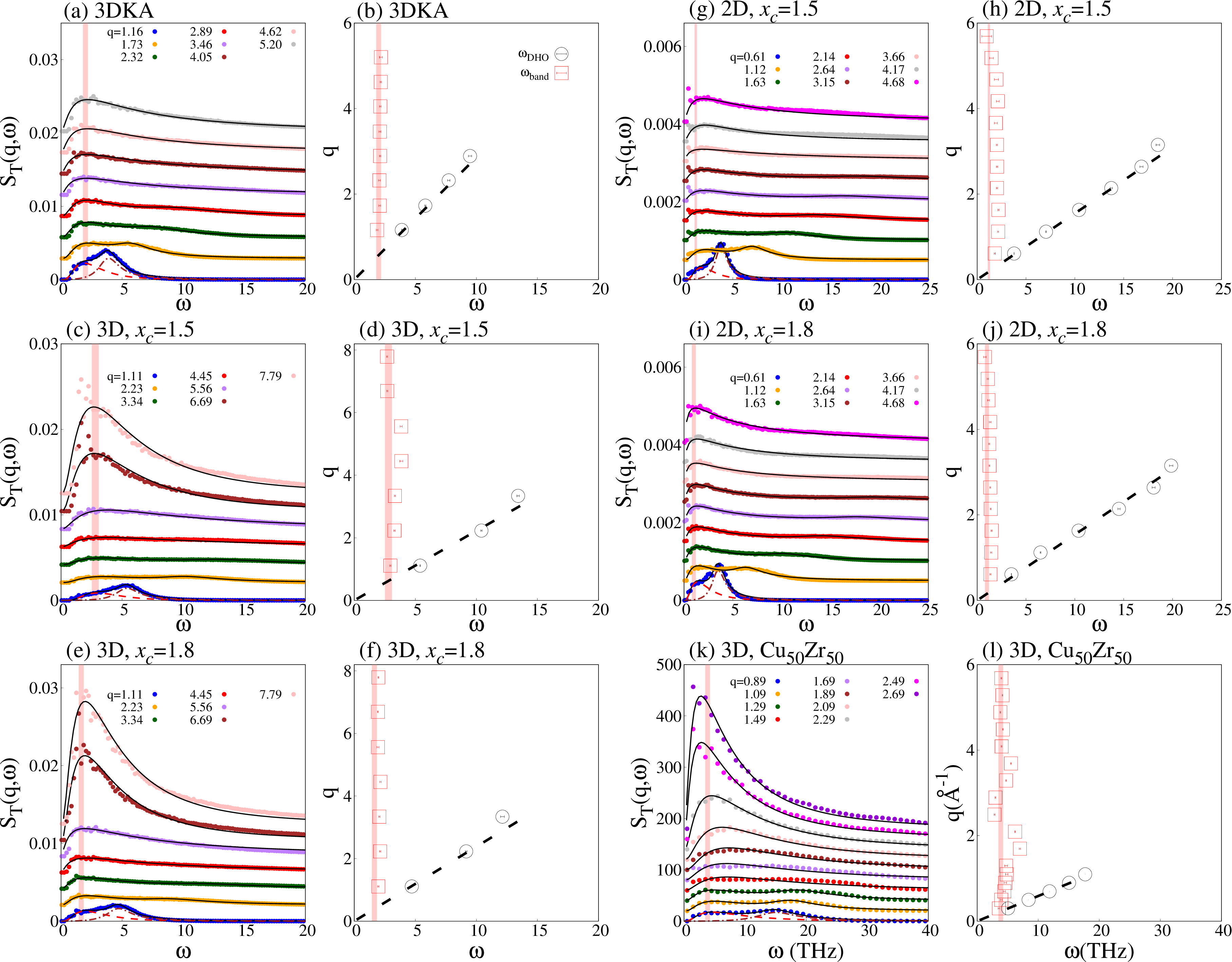}
    \caption{   
Transverse dynamic structure factor $S_T(q,\omega)$ and dispersion relations for six simulated systems, identified by the labels on top of each panel pair. 
Panels are arranged in pairs: the left panel of each pair (a, c, e, g, i, k) shows $S_T(q,\omega)$ as a function of $\omega$ for increasing wave vectors $q$, with superimposed fits (black lines, see text); the right panel 
(b, d, f, h, j, l) shows the corresponding dispersion relations of the acoustic phonon branch $\omega_{\mathrm{DHO}}(q)$ (black circles) and the non-phononic flat band 
$\omega_{\mathrm{band}}(q)$ (red squares) extracted from those fits. 
In all panels, vertical red bands mark the boson peak (BP) frequency independently extracted from the vibrational density of states. 
In the dispersion panels, dashed lines indicate the long-wavelength linear dispersion $\omega = v_T q$, shown only over the range where phonons are well-defined and the DHO contribution can be reliably estimated (see text).
\label{fig:review_numerics}
}
\end{figure*}

In this Section, we provide novel evidence for a flat BP band in numerical models of amorphous solids in two and three dimensions. We consider the following four numerical models, all investigated in square or cubic domains with periodic boundary conditions.
\begin{itemize}
\item[(i)] 
The standard 80:20 KA mixture in 3D at a fixed number density $\rho=1.2$~\cite{kob1995testing,das2022crossover}. We equilibrate the system at $T=2.0$, a temperature above the onset temperature~\cite{sastry2000onset} at which glassy, non-Arrhenius relaxation sets in. The configurations are then energy-minimized with the conjugate-gradient algorithm, and results are averaged over 10 realizations.
\item[(ii)] Polydisperse systems interacting via a family of LJ-like potentials whose attractive tail extends up to a distance modulated by a parameter $x_c$~\cite{dauchotPotential,chattoraj2020role}.
Previous works have demonstrated that decreasing $x_c$ increases the stability of the system
~\cite{chattoraj2020role, gonzalez2020mechanical, gonzalez2020mechanical2, Zheng2021}.
In the following, we consider $x_c = 1.8$ and $x_c = 1.5$, focusing on $N = 8000$ particle systems, in both 2D and 3D.
The density depends on the dimensionality so that the interparticle spacing is fixed, $a_0 = \rho^{-1/D}\simeq 0.977$.
The systems are equilibrated at high temperatures, well above their respective onset temperature, and the energy is then minimized via CJ.
Results are averaged over $50$ realizations in 2D and over 20 in 3D.
\item[(iii)] A model three-dimensional $\mathrm{Cu_{50}Zr_{50}}$ metallic glass \cite{10.1063/1.5131500}. An empirical potential function is used to simulate the interactions in a realistic three-dimensional $\mathrm{Cu_{50}Zr_{50}}$ metallic glass sample, which is represented by a simulation box of $14000$ atoms.
We equilibrate a metallic liquid for $2$ ns at $1800$ K, the time duration of which is much longer than the $\alpha$ relaxation time, and then quench the liquid to $300$ K with a constant cooling rate of $10^{12}$ K/s. These simulations are performed in the NPT ensemble at zero pressure~\cite{10.1063/1.447334}. The energy of the resulting configuration is then minimized via CJ at constant volume based on the last step of cooling.
\end{itemize}
We diagonalize the Hessian matrix of each inherent state produced by the energy minimization to obtain $dN$ eigenvalues $\omega^2$ and eigenvectors.
We estimate the BP frequency $\omega_{\text{BP}}$ from the peak of the normalized density of states, $g(\omega)/\omega^{D-1}$, and evaluate the dynamical structure factors from the normal mode analysis, as outlined in Ref. ~\cite{tanakaPRR}. 

Figure~\ref{fig:review_numerics} displays the dynamical transverse structure factor 
$S_T(q,\omega)$ for the different systems (see Fig.~S1 in the Supporting Information 
(SI) for the corresponding density plots). A dispersionless peak is observed at a 
frequency matching that of the boson peak (vertical red bands). This feature is robust 
with respect to both the interaction potential and the dimensionality. Similar behavior 
is also found in the longitudinal component $S_L(q,\omega)$ (Fig.~S2 in the SI).

To provide a more quantitative analysis, we fit $S_T(q,\omega)$ to a two-component 
model. The first component is a log-normal function accounting for the non-phononic 
flat-band contribution,
\begin{equation}
    S_{\mathrm{nph}}(q,\omega) \propto \frac{1}{\omega \sigma \sqrt{2\pi}}
    \exp \left(-\frac{\left(\ln \omega -\mu\right)^2}{2 \sigma^2}\right)
    \label{lognormal}
\end{equation}
The second component is a damped harmonic oscillator (DHO) term describing the 
phononic contribution,
\begin{equation}
    S_{\mathrm{DHO}}(q,\omega)\propto 
    \frac{\omega^2\,\Gamma(q)}
    {\left(\omega^2-\omega^2_{\mathrm{DHO}}(q)\right)^2+\omega^2 \Gamma^2(q)}.
    \label{DHO}
\end{equation}
At small $q$, both components contribute to the spectrum and are fitted simultaneously. 
As $q$ increases, phonons become progressively damped and the DHO contribution to 
the total spectral weight diminishes. In this regime the DHO parameters can no longer 
be reliably estimated, and we accordingly drop the DHO term, retaining only 
$S_{\mathrm{nph}}$. Operatively, we discard the DHO whenever the fitted 
$\omega_{\mathrm{DHO}}$ deviates by more than $10\%$ from the phonon frequency 
expected from the acoustic dispersion.

From the fits (black curves in Fig.~\ref{fig:review_numerics}), we extract the acoustic 
phonon dispersion $\omega_{\mathrm{DHO}}(q)$ and the flat-band frequency 
$\omega_{\mathrm{band}}(q)$, defined as the peak of the log-normal, 
$\omega_{\mathrm{band}} = \exp(\mu - \sigma^2)$. 
Both quantities are reported in  Fig.~\ref{fig:review_numerics} for all simulated models. The flat-band signal (red empty squares) is largely independent of $q$, demonstrating the dispersionless 
character of the non-phononic modes. Its frequency is in good agreement with the 
boson peak (BP) frequency, indicated by the vertical red band and independently 
extracted from the vibrational density of states. This confirms that 
$\omega_{\mathrm{band}} \simeq \omega_{\mathrm{BP}}$ across all systems considered.

\section*{Experimental evidence}
After presenting substantial evidence for the universality of the flat BP band in simulated glasses, we now turn to the question of whether experimental evidence supporting its existence can be found. Since many of the presented figures are taken from the existing literature, we will use the symbols $q$, and $k$ interchangeably to indicate the wavevector.

\subsection*{Granular packing}
\begin{figure}[!t]
    \centering
    \includegraphics[width=0.85\linewidth]{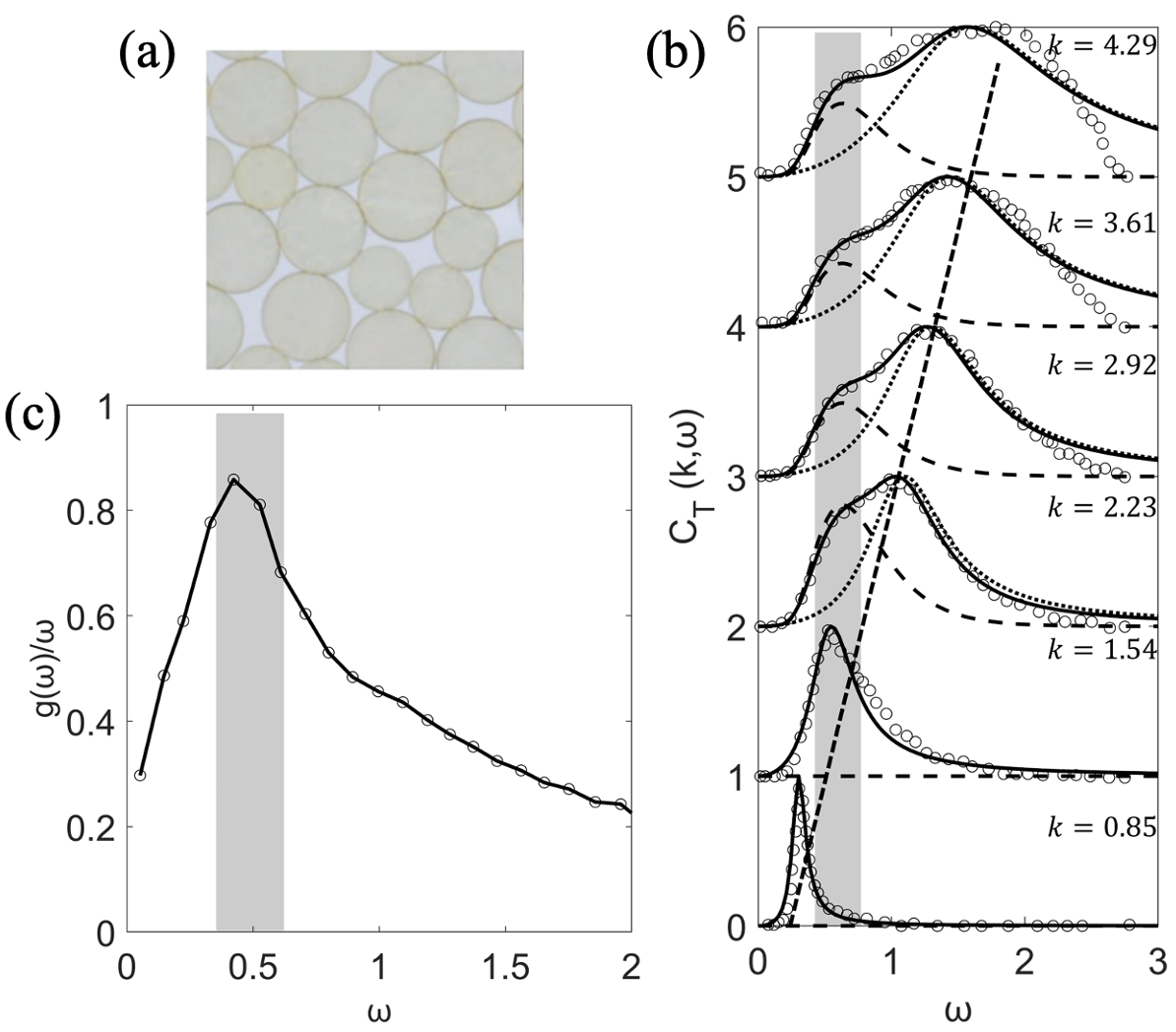}
    \caption{\textbf{(a)} An image of the experimental system in Ref.~\cite{PhysRevB.98.174207} showing the random packing of bi-disperse photoelastic disks. \textbf{(b)} The transverse correlation function as a function of the frequency $\omega$ for different values of the wavevector $k$. Solid lines are the fits combining the DHO phononic contribution \eqref{DHO} and the non-phononic one \eqref{lognormal}. The two contributions are shown independently with dotted and dashed lines. The vertical grey bar indicates the BP frequency. \textbf{(c)} The reduced density of states showing the emergence of a BP anomaly around $\omega_{\text{BP}}\approx 0.5$ (vertical grey bar). Data are taken with permission of the authors in Ref.~\cite{PhysRevB.98.174207}.}
    \label{figyinqiao}
\end{figure}

We begin by revisiting the experimental data and theoretical interpretation presented by some of us in Refs.~\cite{PhysRevB.98.174207} and~\cite{PhysRevLett.133.188302}. The experimental setup of Ref.~\cite{PhysRevB.98.174207} consists of a two-dimensional jammed packing of bidisperse photoelastic disks (see Fig.~\ref{figyinqiao}(a)), in which interparticle forces can be measured with high precision. By employing advanced image processing and force measurement techniques (see also Refs.~\cite{Majmudar2005,Zhang2017}), the harmonic Hessian matrix was reconstructed. Using standard diagonalization methods, the authors were then able to extract the vibrational density of states, as well as the transverse and longitudinal current correlation functions $C_\alpha(k,\omega)$, which are directly related to the dynamical structure factors, $S_\alpha(q,\omega)=\frac{q^2}{\omega^2}C_\alpha(q,\omega)$.

In Ref.~\cite{PhysRevB.98.174207}, the current correlation functions were modeled using a single phononic contribution, corresponding to longitudinal and transverse acoustic phonons. 
While this approximation appears adequate for the longitudinal component (see the blue curves in Fig.~3 of Ref.~\cite{PhysRevB.98.174207}), the interpretation becomes more delicate for the transverse part, which we reproduce in
Fig.~\ref{figyinqiao}(b).
Indeed, only at low wavevectors, specifically $k = 0.85$ and $1.54$, $C_T(k,\omega)$ exhibits a peak which can be accurately fitted by a damped harmonic oscillator (DHO) form, \eqref{DHO}.

For large values of the wavevector $k$, a pronounced low-frequency shoulder clearly emerges, indicating the presence of an additional contribution that was not accounted for in Ref.~\cite{PhysRevB.98.174207}.
This shoulder becomes increasingly pronounced with growing $k$. 
Following Ref.~\cite{tanakaNatPhys}, we model this non-phononic contribution using a $k$-independent log-normal function, \eqref{lognormal}.

As shown in Fig.~\ref{figyinqiao}(b), modeling the total correlation function as the sum of a DHO term [\eqref{DHO}] and a non-dispersive term [\eqref{lognormal}] yields excellent fits across the entire $k$ range- fits that would be otherwise unattainable. We do not attribute any physical significance to the adopted log-normal functional form, and have verified that alternative ones can also describe the data.

Importantly, the characteristic energy scale of the non-dispersive mode described by \eqref{lognormal} matches well with the boson peak frequency extracted from the reduced vibrational density of states (VDOS), as shown in Fig.~\ref{figyinqiao}(c) and highlighted by a vertical grey bar in both panels.

\begin{figure}[!t]
    \centering
    \includegraphics[width=0.85\linewidth]{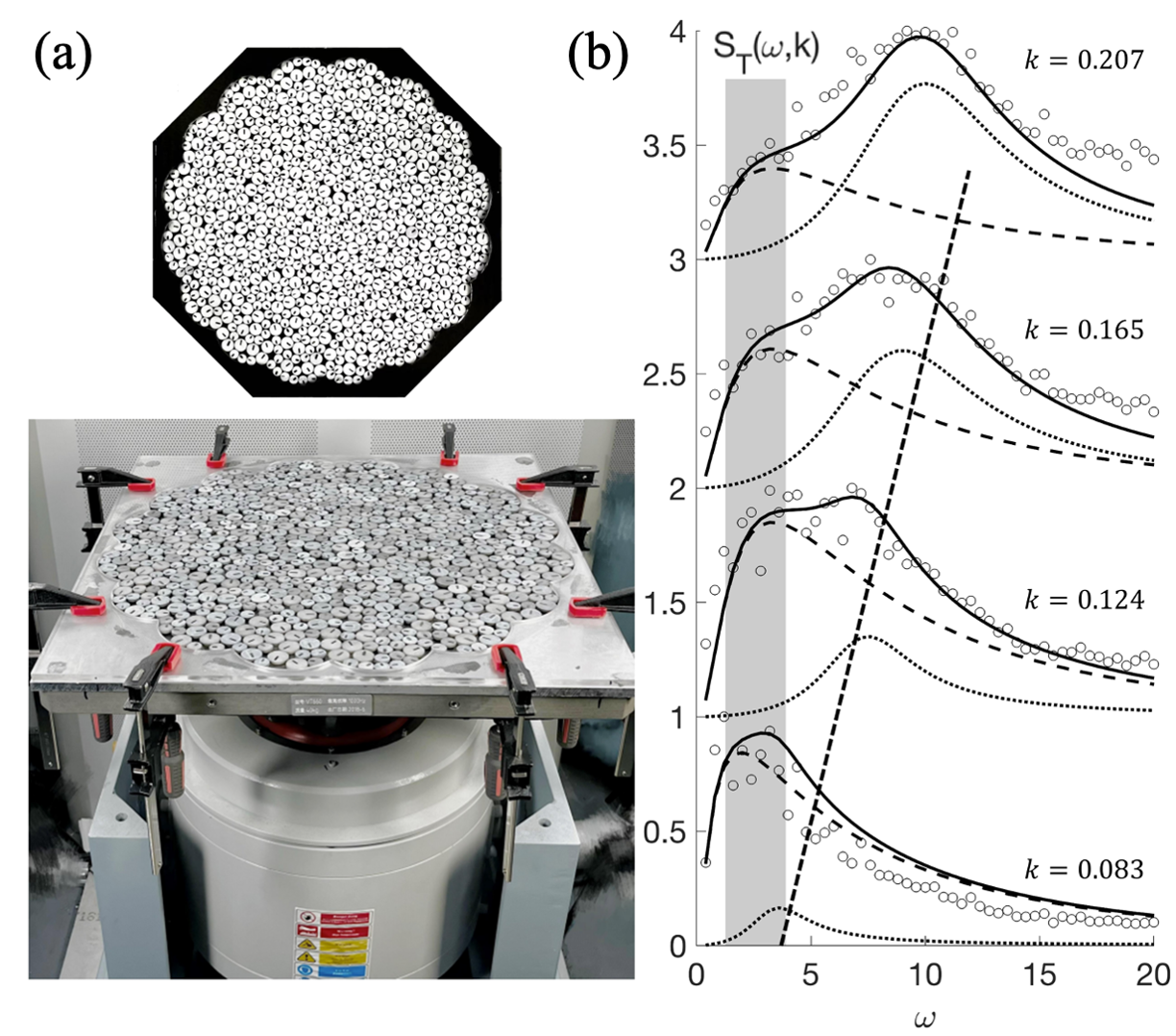}
    \caption{\textbf{(a)} Images of the experimental setup considered in Ref.~\cite{PhysRevLett.133.188302}. \textbf{(b)} The experimental dynamical structure factor $S_T(\omega,k)$. Open symbols are the experimental data. Dotted and dashed lines are the phononic and non-phononic contributions, Eqs.~\ref{DHO}-\ref{lognormal}. The solid lines are the fit involving the sum of both of them. The vertical grey bar indicates the position of the BP independently estimated in Ref.~\cite{PhysRevLett.133.188302} using the reduced density of states.}
    \label{figactive}
\end{figure}

Ref.~\cite{PhysRevLett.133.188302} investigated a related, yet distinct, two-dimensional experimental system composed of bi-disperse granular particles vertically agitated by an electromagnetic shaker (see Fig.~\ref{figactive}(a)). 
By tracking the position of each particle with a CCD camera, the full dynamics, including the displacement field, could be reconstructed. 
From this, the displacement correlations were computed and subsequently used to derive the dynamical matrix. 
Diagonalizing this matrix provided direct access to both the vibrational density of states and the dynamical structure factor $S(\omega,k)$.

In Fig.~\ref{figactive}(b), we reproduce the experimental data for $S_T(\omega,k)$ as a function of frequency for various values of $k$. 
Also in this case, $S_T(\omega,k)$ can only be well captured at all wavevectors if modeled as the sum of a 
damped harmonic oscillator term, associated with the transverse acoustic phonon, and a dispersionless component, 
with a frequency scale which aligns well with the boson peak frequency extracted from the vibrational density of states (see Ref.~\cite{PhysRevLett.133.188302} for details). Importantly, in a mono-disperse crystalline sample, where the BP was clearly absent, no trace of this dispersionless band was reported.

Overall, these revised analyses reveal that the experimental data from Refs.~\cite{PhysRevB.98.174207, PhysRevLett.133.188302} already contain compelling evidence for the presence of a flat band, which aligns precisely with the BP frequency. This convergence underscores the universality of the BP flat band and reinforces the validity of the numerical analyses.

\begin{figure}[!t]
    \centering
    \includegraphics[width=0.55\linewidth]{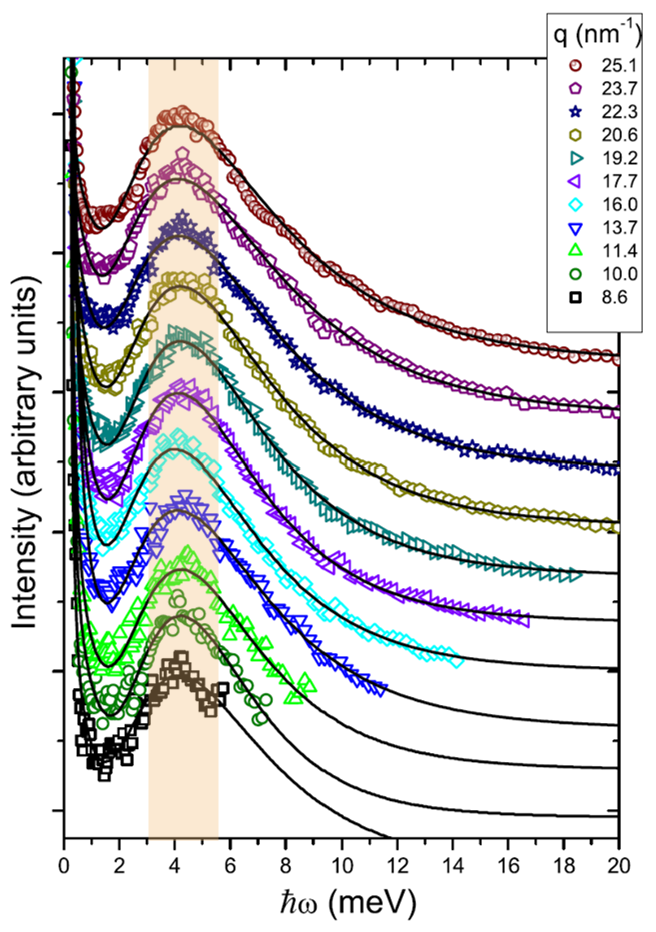}
    \caption{Inelastic neutron scattering spectra of vitreous silica at room temperature and different wavevectors. The dispersionless peak is indicated with a shaded vertical color band and its energy corresponds well with the BP frequency $\omega_{\text{BP}}\approx 4.2$ meV. Figure adapted with permission from \cite{PhysRevB.77.214309}.}
    \label{fig:baldi}
\end{figure}

\subsection*{Vitreous silica}

\paragraph*{Bulk vitreous silica}--
Refs.~\cite{Ruzicka04,PhysRevB.77.214309} conducted a detailed experimental study of the dynamic structure factor in vitreous silica using both inelastic neutron scattering and inelastic x-ray scattering. See also \cite{BaldiChapter}.
Remarkably, both techniques revealed a low-frequency, non-dispersive component around the boson peak frequency, $\omega_{\text{BP}} \approx 4.2$ meV, loosely labeled as ``\textit{transverse-optic}''. Figure~\ref{fig:baldi} reproduces the INS data from Ref.~\cite{PhysRevB.77.214309}, where this feature is particularly evident. The non-dispersive peak emerges only for $q > 4$ nm$^{-1}$ and appears to influence both the longitudinal sound dispersion and the damping of sound waves. In our view, this mode corresponds to the flat boson-peak band.

\begin{figure}[!t]
    \centering
    \includegraphics[width=0.75\linewidth]{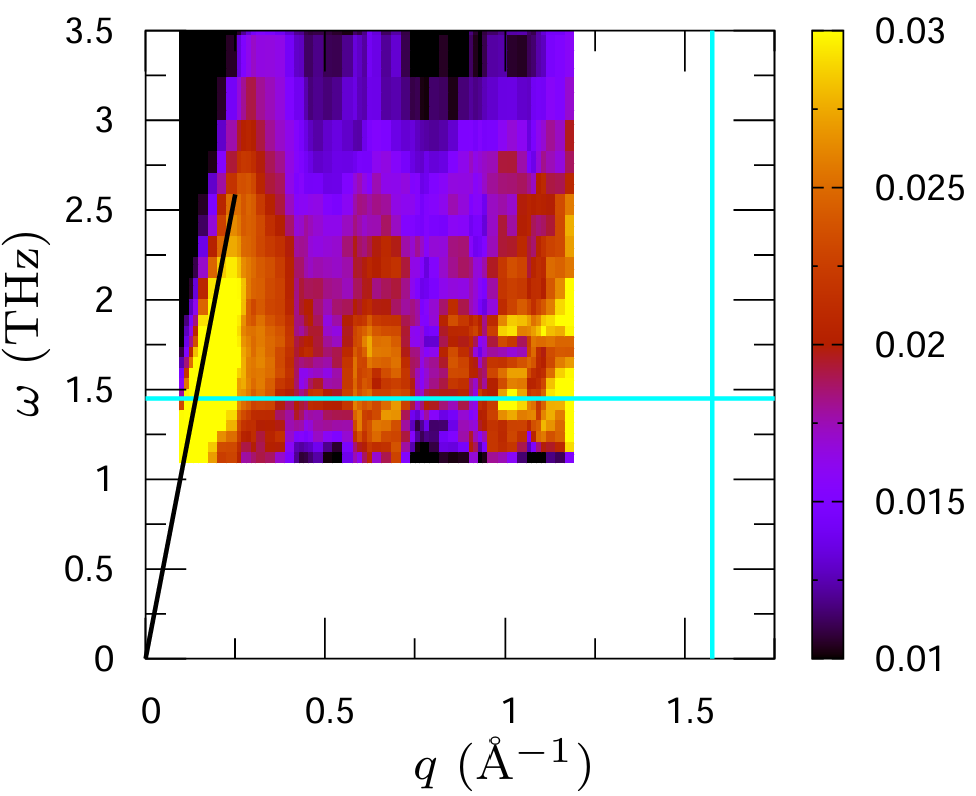}
    \caption{Longitudinal dynamical structure factor of silica glass measured via inelastic X-ray scattering. Figure taken from \cite{mizuno2025bosonpeakcovalentnetwork} with permission from the authors. The solid black line is the dispersion of the longitudinal acoustic phonon. The horizontal cyan solid line indicates the BP frequency.}
\label{newexp}
\end{figure}

This perspective has been recently reinforced by Ref.~\cite{mizuno2025bosonpeakcovalentnetwork}, which presents experimental data for the longitudinal dynamic structure factor of silica glass (reproduced in Fig.\ref{newexp}). The analysis supports the existence of a non-phononic dispersionless band at the boson peak frequency. Remarkably, in the case of silica glass, this band is observed not only in the transverse but also in the longitudinal spectrum. We observe similar features in the amorphous CuZr and KA simulated models.

\paragraph*{Monolayer 2D silica}-- 
More recently, inelastic helium-atom scattering experiments~\cite{Tomterud2023} revealed the presence of a boson peak in two-dimensional silica. 
In this technique, the measured differential reflection coefficient is directly related to the dynamical structure factor, and hence to the spectral function (the imaginary part of the Green's function), which encodes the system's vibrational modes.

\begin{figure}[!t]
    \centering
    \includegraphics[width=\linewidth]{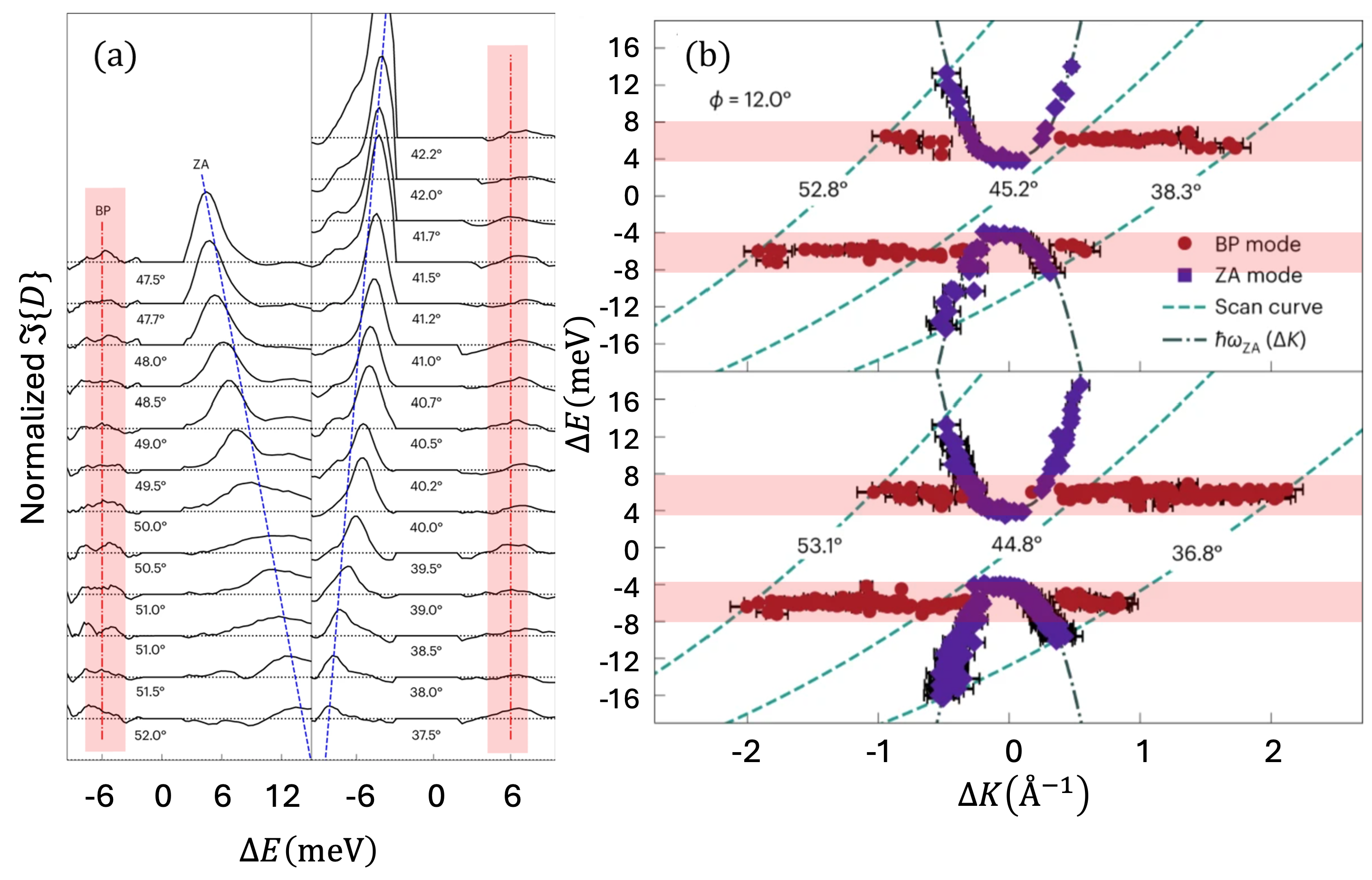}
    \caption{\textbf{(a)} Measured spectral function normalized by $\omega$ as a function of the energy transfer $\Delta E$. The red bars highlight the position of the BP, that is around $6$ meV. \textbf{(b)} The dispersion relation of the lowest excitations in the system as a function of the parallel momentum transfer $\Delta K$. A dispersionless BP band is evident and its energy perfectly matches the location of the BP in panel (a). Figures adapted with permission from \cite{Tomterud2023}.}
    \label{fig:exp2d}
\end{figure}

Figure~\ref{fig:exp2d}(a) shows the normalized spectral function as a function of energy transfer $\Delta E$ for various polar incident angles. 
A strong peak follows the dispersion of the flexural (ZA) phonon, while an additional weaker peak appears near $\Delta E = \pm 6$ meV (red bar), marking the boson peak anomaly. This feature becomes more evident when spectra are averaged over all angles (see Fig. 3 in \cite{Tomterud2023}).

To investigate the origin of the BP anomaly, the dispersion relation of the lowest-energy modes was extracted from the same data, as shown in Fig.~\ref{fig:exp2d}(b). 
The ZA mode follows the expected gapped, quadratic dispersion \cite{PhysRevB.88.115418} (purple symbols). 
In contrast, the so-called BP band (red symbols) exhibits a flat dispersion, largely independent of the parallel momentum transfer $\Delta K$. Its energy closely matches the BP feature identified in panel (a), reinforcing the interpretation that the BP anomaly originates from a dispersionless non-phononic band.

\begin{figure}[!t]
    \centering
    \includegraphics[width=0.8\linewidth]{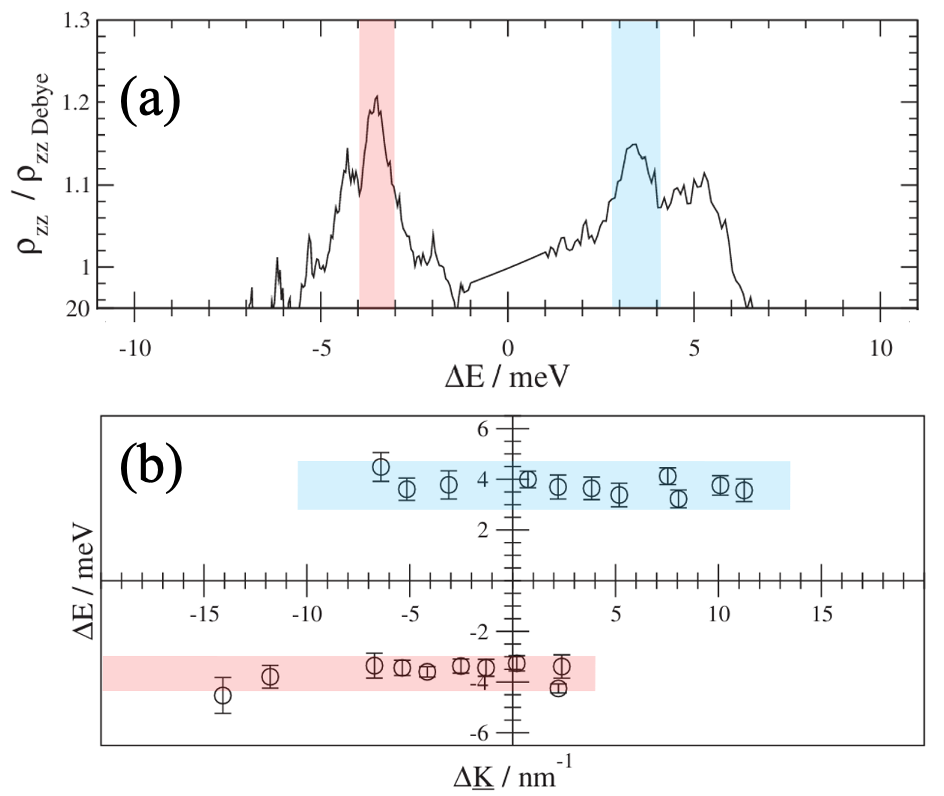}
    \caption{\textbf{(a)} Measured spectral density divided by
Debye limit. \textbf{(b)} Energy dispersion of the dispersionless BP band as extracted from the time-of-flight spectrum (see Fig. (1) in \cite{PhysRevLett.99.035503}). The color bars highlight the matching of the BP frequency with the energy of the flat band. Figures reproduced from the arXiv preprint: \href{https://arxiv.org/abs/2205.02486}{2205.02486}.}
    \label{fig:expsurf}
\end{figure}

\paragraph*{2D $v$-SiO$_2$}--
Ref.~\cite{PhysRevLett.99.035503} employed a similar experimental technique to reveal the emergence of a boson peak in 2D vitreous silica deposited on a substrate. 
As noted in the abstract, ``\textit{The boson peak appears as a dispersionless mode of $\approx 4$ meV in the recorded time-of-flight spectra}.'' 
Figure~\ref{fig:expsurf}(a) shows the experimental spectral function normalized by the two-dimensional Debye prediction, where the BP anomaly is clearly visible and highlighted with colored bars. 
Panel (b) presents the energy of the non-phononic band extracted from the TOF spectra (see Fig. 1 in \cite{PhysRevLett.99.035503}), which again exhibits a flat dispersion closely matching the BP feature in the spectral function.

\paragraph*{Bilayer silica}--
Ref.~\cite{PhysRevB.111.205423} recently reported the observation of a boson peak in bilayer silica using helium atom scattering. 
The data reveal a flat, dispersionless BP band whose energy coincides with the anomaly observed in the reduced vibrational density of states, confirming the presence of a characteristic BP band in this bilayer system.


\subsection*{Amorphous metallic systems}
\begin{figure}[!t]
    \centering
    \includegraphics[width=0.85\linewidth]{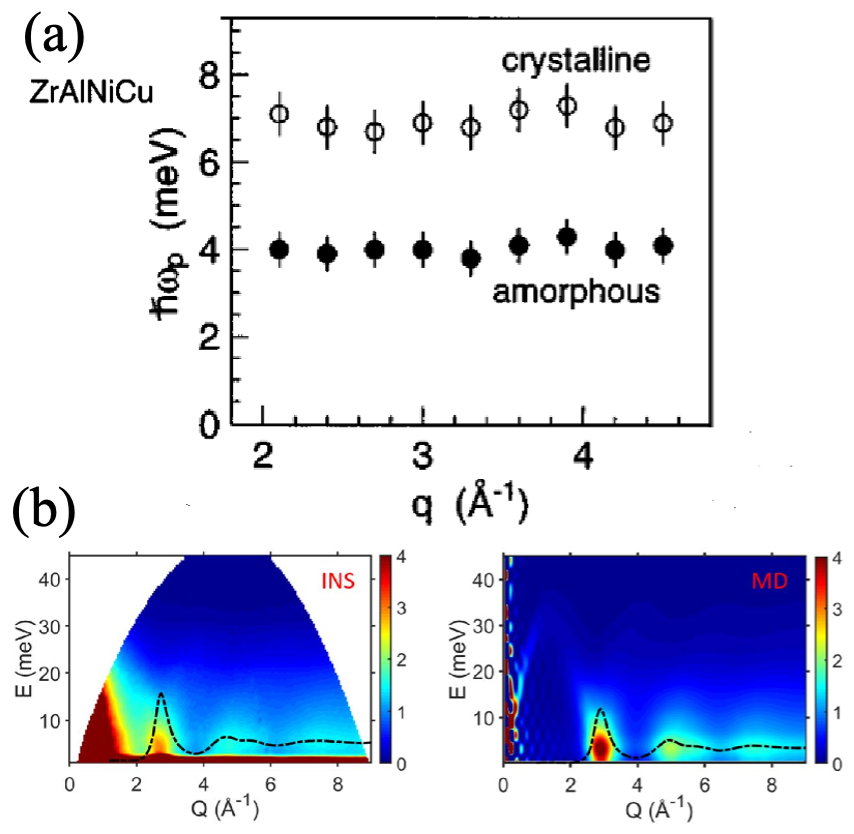}
    \caption{\textbf{(a)} Frequency $\omega_p$ of the of the inelastic maximum of $S(q,\omega)$, measured via neutron scattering, as function of $q$ in of ZrAlNiCu alloy.
    \textbf{(b)} INS measurement (left) and simulations (right) of the Zr$_{46}$Cu$_{46}$Al$_8$ density of states normalized by $E^2$.  The data show a dispersionless BP around 5.5 meV. The static structure factor S(Q) is superimposed.
    Adapted with permissions from  \cite{PhysRevB.53.12107} (panel a) and \cite{Li2025} (panel b).}
    \label{fig:expmetal}
\end{figure}

\paragraph*{ZrTiCuNiBe and ZrAlNiCu}--
In Ref.~\cite{PhysRevB.53.12107}, coherent neutron scattering measurements spanning five decades in energy were performed on the amorphous alloys ZrTiCuNiBe and ZrAlNiCu. The vibrational spectra clearly displayed a boson peak. Importantly, the BP position was largely non-dispersive with respect to wavevector, while its intensity was strongly correlated with the static structure factor. Figure~\ref{fig:expmetal}(a) shows the data for ZrAlNiCu, confirming that the mode responsible for the BP was approximately insensitive to the wavevector $q$, i.e., it was dispersionless.

\paragraph*{Zr$_{46}$Cu$_{46}$Al$_8$}--
Recent studies using inelastic neutron scattering on Zr$_{46}$Cu$_{46}$Al$_8$ metallic glass~\cite{Li2025} reported similar observations, as illustrated in Fig~\ref{fig:expmetal}(b). Within the instrumental resolution, the BP energy remained nearly constant across wavevectors, showing negligible dispersion. Moreover, the energy of the flat band closely matched the BP energy measured independently via Raman scattering and heat-capacity experiments. Complementary molecular dynamics simulations of Zr-Cu-Al metallic glasses corroborated these findings, further supporting the interpretation of the BP as originated from a dispersionless non-phononic band.

\subsection*{Polymers}
\begin{figure}[!t]
    \centering
    \includegraphics[width=0.65\linewidth]{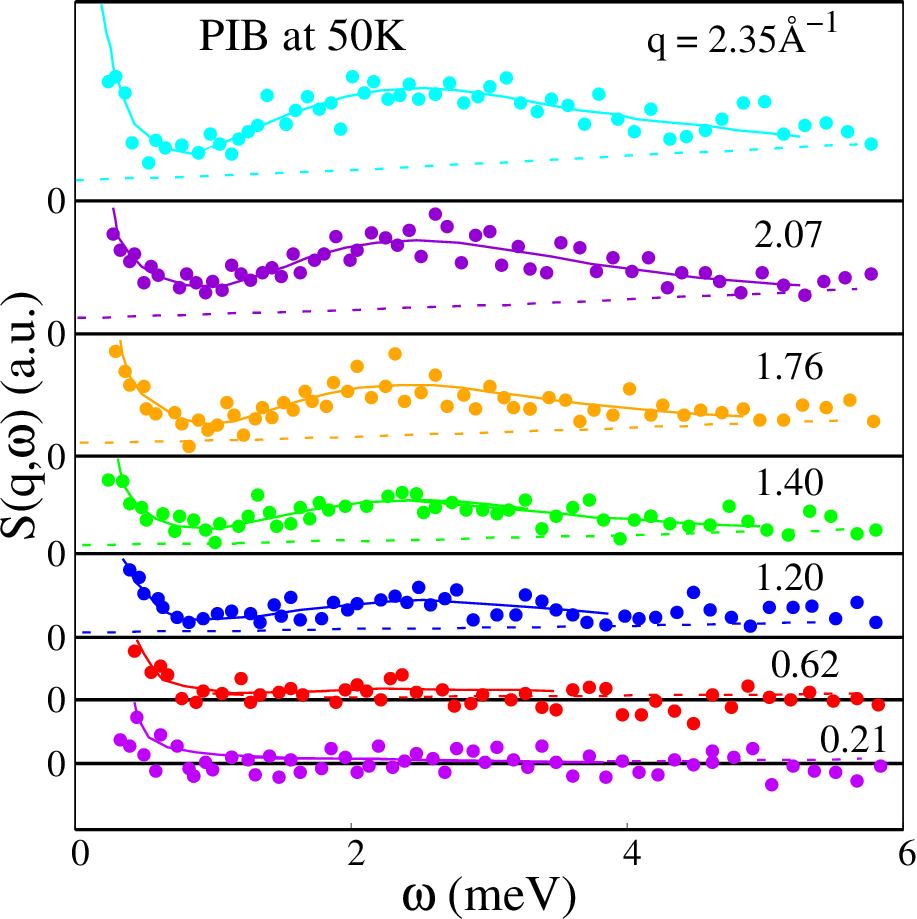}
    \caption{Inelastic neutron-scattering spectra of PIB at $T=50$ K as a function of the wavevector $q$. The dashed lines indicate the contribution of the sound wave. The solid line is a fit of the dispersionless low-frequency peak. 
    Data taken from Ref.~\cite{inoue1991low}.}
    \label{fig:poly}
\end{figure}

In Ref.~\cite{inoue1991low}, inelastic neutron-scattering measurements were performed on various organic amorphous polymers. Figure~\ref{fig:poly} reproduces the experimental results for polyisobutylene (PIB) at 50 K. An non-phononic low-energy signal was observed at approximately 2.6 meV in $S(q,\omega)$. The position of this peak was independent of the wavevector $q$, while its intensity increased with $q$ without any noticeable change in shape. Notably, the energy of this band was very close to the boson peak anomaly independently observed in the vibrational density of states around 2 meV. These findings provide further support for the interpretation that a nonphononic flat band is responsible for the boson peak.

\subsection*{Other examples}
In Ref.~\cite{PhysRevLett.96.045502}, the spectrum of a lithium diborate glass, Li$_2$O$_2$B$_2$O$_3$, was measured using Brillouin and x-ray scattering techniques. Although the main focus of that work was the nature of sound-wave damping in glasses, the authors also reported a dispersionless excitation. By comparing spectra at wave vectors of $23.4$, $26.4$, and $29.4$ nm$^{-1}$, they found that the energy of this signal was independent of the wave vector and matched well the boson peak frequency. These observations are consistent with our interpretation that, in this system as well, the BP originates from a flat, dispersionless band.

\begin{figure}[!t]
    \centering
    \includegraphics[width=0.7\linewidth]{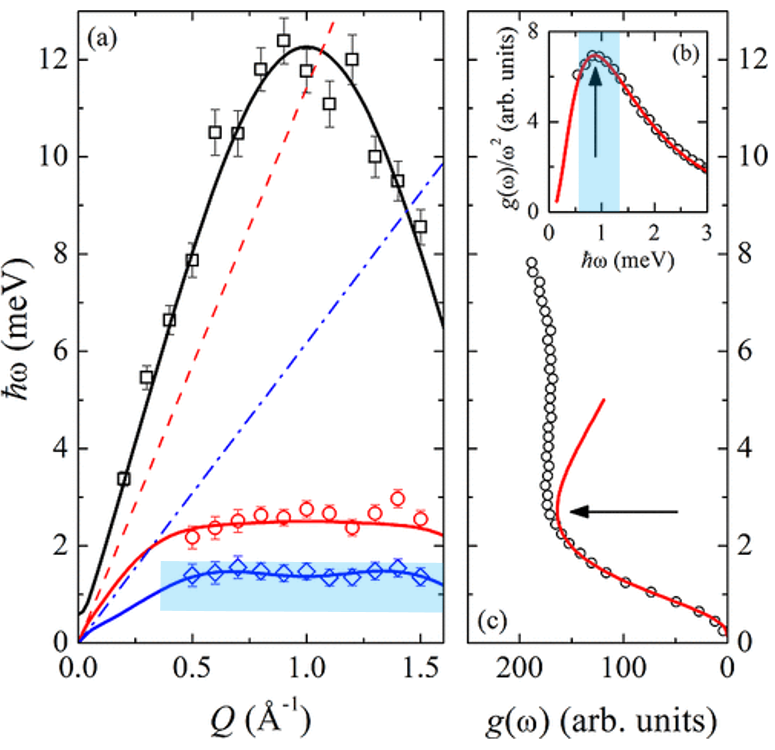}
    \caption{\textbf{(a)} Inelastic neutron scattering data (symbols) of the dispersion of low-energy excitations in glassy SiSe$_2$. Solid lines are fits to the data. Dashed and dash-dotted lines are the linear dispersions obtained using the estimated speeds of sound. Red and blue symbols show two dispersionless flat bands. The color band highlights the BP region evident in panel (b). \textbf{(b)} Reduced density of states $g(\omega)/\omega^2$. The arrow indicates the position of the BP. \textbf{(c)} Density of states $g(\omega)$. Figure adapted with permission from Ref.~\cite{Zanatta2013}.}
    \label{fig00}
\end{figure}

Finally, we note that Ref.~\cite{BaldiChapter} (see Section 3.3) already provided a comprehensive summary of experimental evidence supporting this scenario, updated through 2022. The list spans a broad range of amorphous materials, including v-GeO$_2$ glass \cite{Bove_2005}, metallic glasses \cite{10.1063/1.3640002,PhysRevB.78.052202}, chalcogenides \cite{PhysRevB.82.115201,Zanatta2013}, organic systems \cite{PhysRevB.85.134204,10.1063/1.4998696,Scopigno_2003}, and even water \cite{PhysRevE.71.011501}, thus enlarging the spectrum of cases where the flat band has been observed. In Fig.~\ref{fig00}, we reproduce the results of Ref.~\cite{Zanatta2013} for glassy SiSe$_2$, where inelastic neutron scattering clearly reveals a dispersionless flat band whose energy coincides with the BP frequency.

\section*{Linking Structure and Dynamics at the Boson Peak}
We investigate the spatial characteristics of the vibrational modes responsible for the flat-band signal in $S_\alpha(q,\omega)$ by analyzing the reduced spectrum $B_\alpha(q,\omega)$ (see \eqref{eq:reducedSq}). This construction removes the standard phononic contributions and follows earlier approaches~\cite{PhysRevB.44.6746,PhysRevB.62.3181,Schober_2004} as well as more recent work~\cite{Li2025}. We focus here on the transverse component ($\alpha = T$); results for the longitudinal case are presented in the SI.

\begin{figure}[!t]
    \centering
    \includegraphics[width=0.85\linewidth]{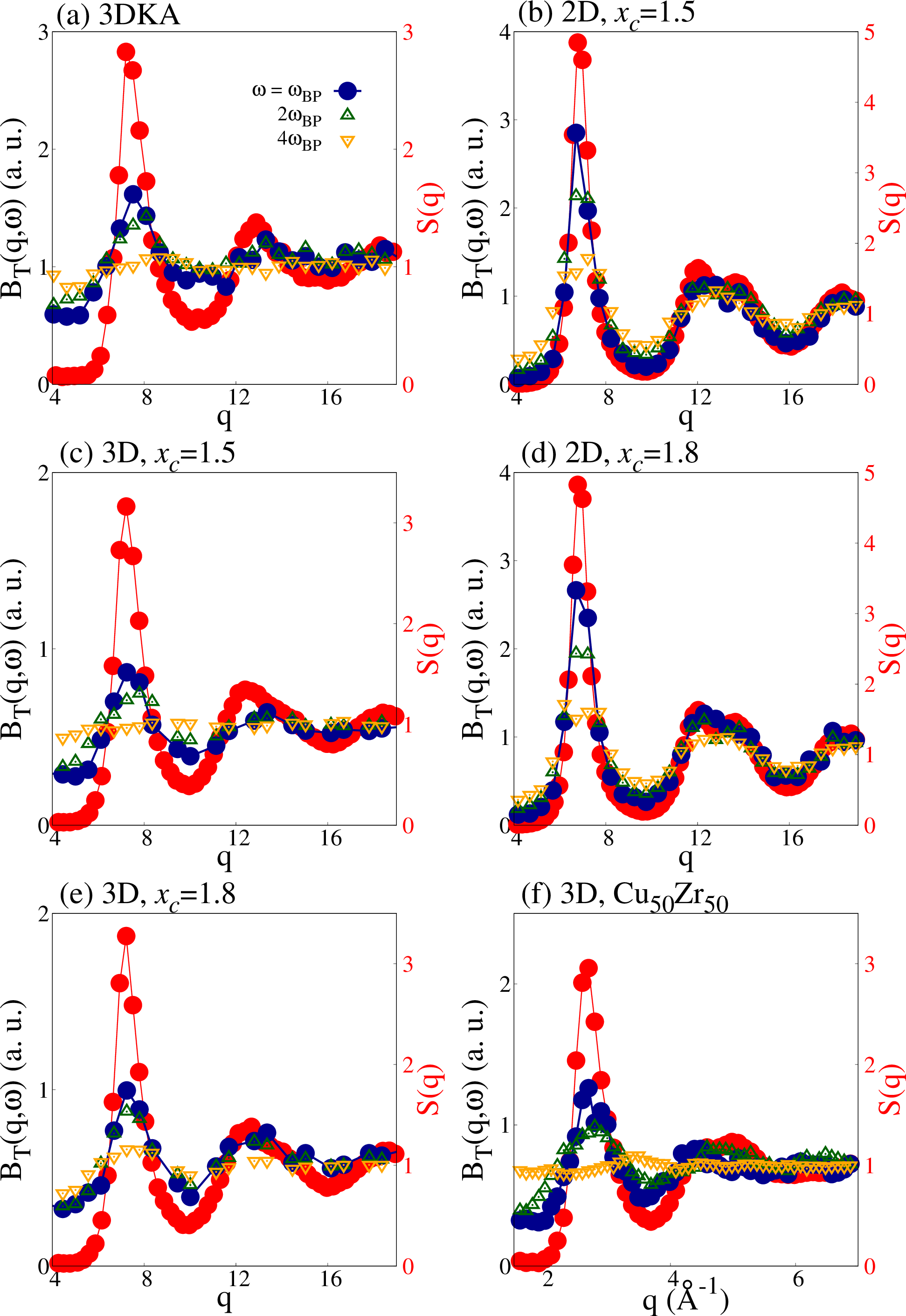}
    \caption{Comparison between the reduced transverse dynamical structure factor $B_T(q,\omega)$ calculated at different frequencies and the static structure factor $S(q)$ for different simulated systems. See SI Fig.~S2 for a comparison between $B_L(q,\omega)$ and $S(q)$.}
    \label{corr}      
\end{figure}

Fig.~\ref{corr} shows $B_T(\omega,q)$ at selected frequencies for all model systems considered in Fig.~\ref{fig:review_numerics}. We observe a close correspondence between $B_T(\omega,q)$ evaluated at the boson peak frequency (blue) and the static structure factor $S(q)$ (red). This correlation is particularly pronounced in two dimensions, while remaining qualitatively valid in three dimensions. It indicates that the excess modes giving rise to the flat band exhibit a spatial organization that is strongly correlated with underlying density fluctuations. This observation supports a structural interpretation of the boson peak and is consistent with recent real-time measurements of medium-range-order fluctuations in glasses~\cite{tian2021structural}.

Because $S(q)$ vanishes at low wave vectors, this structural correspondence implies that the flat-band intensity must intrinsically drop to zero as $q\!\to\!0$, explaining why the signal is not observed in this regime.
This settles the question of whether low-$q$ flat-band features are masked by acoustic phonons: their suppression is a genuine physical property of the excess modes themselves and possibly their coupling to acoustic phonons.

\section*{Challenging observations}
Motivated by the well-known principle that falsifying a scientific hypothesis is often easier than confirming it, we examined the existing literature for counterexamples. 

One counterexample we identified is Ref.~\cite{MonacoMossa}, which reports a molecular dynamics investigation of a monodisperse Lennard-Jones glass. 
In that work, no flat, dispersionless band at the BP frequency was observed (see Fig. 1 therein). 
This work focused on the vibrational properties at mesoscopic length scales, and hence considered very small wavevectors, the largest one being approximately $1/10$ of the first diffraction peak.
As discussed in the previous paragraph, this wavevector falls well below the threshold for detecting the flat-band signal. This provides a simple and natural explanation for the absence of a flat band in the experimental measurements reported in Ref.~\cite{MonacoMossa}.

Another counterexample is in the early investigation~\cite{PhysRevLett.50.49} of the dynamic structure factor in a Mg$_{70}$Zn$_{30}$ metallic glass, which reported a distinct peak with a clear dispersive character (see Fig. 2 in Ref.~\cite{PhysRevLett.50.49}), with its energy oscillating between $4$ and $6$ meV. Further studies are required to determine whether this constitutes a genuine counterexample to the dispersionless nature of the BP, or at least challenges its universality.

\section*{Constraints on theoretical models}
\label{sec:constraints}
The prevalent observation of a non-phononic flat band in the dynamical structure factor at the boson peak frequency places strong empirical constraints on any microscopic theory of the boson peak (see Ref.~\cite{ramos2022low} for a review).  
Specifically, a successful framework should (i) account for the emergence of a non-phononic, weakly dispersive band distinct from acoustic phonons; (ii) explain why the energy associated with this band coincides with the excess in the vibrational density of states; and (iii) reproduce the observation that its intensity inherits the $q$-dependence of the static structure factor
$S(q)$.
To our knowledge, no existing theoretical framework simultaneously accounts for all three aspects in its current formulation.  

As summarized in Table~\ref{nono}, the flat-band phenomenology already provides a sharp diagnostic.
Theoretical approaches that interpret the boson peak as an exclusive phonon-based feature, arising from a modified, scattered, or damped acoustic branch, cannot naturally account for the emergence of a distinct, weakly dispersive band in the dynamical structure factor.
By contrast, frameworks that posit vibrational excitations beyond acoustic phonons, and coupled to them, remain, at least in principle, compatible with the full set of constraints.

\begin{table*}[t]
\centering
\begin{tabular}{|c|c|}
\hline
\multicolumn{2}{|c|}{\textbf{Flat BP band: a model-discriminating observation}} \\ \hline
\textbf{Compatible models} & \textbf{Models under tension} \\ \hline

Quasi-localised vibrations \cite{SCHOBER2011501} &
Heterogeneous elasticity \cite{doi:10.1142/9781800612587_0009} \\

Soft-potential / QLMs \cite{Galperin89,schober1996low,Gurevich03,Parshin07,Ramos1998,10.1063/5.0147889} &
Van Hove smoothening \cite{PhysRevLett.112.025502} \\

Stringlet excitations \cite{10.1063/5.0039162,tanakaNatPhys,tanakaPRR,10.1063/5.0210057,Jiang_2024} &
Damping-based models \cite{PhysRevLett.122.145501,PhysRevResearch.2.013267,Ding2025} \\

Isostaticity / marginal stability \cite{C4SM00561A,mizuno2025bosonpeakcovalentnetwork} &
\dots \\

Generalized HET \cite{Schirmacher2024} &
\dots \\ \hline
\end{tabular}
\caption{Non-exhaustive list of theoretical models that are compatible or in tension with the observation of a flat BP band. QLMs stands for quasilocalised modes and HET for heterogeneous elasticity.
\label{nono}}
\end{table*}

Determining which existing or new theoretical framework most accurately describes the boson peak is closely connected to understanding the microscopic nature of the flat BP band. 
Once a model successfully reproduces the criteria outlined above, several fundamental questions remain: What is the microscopic origin of this flat band? Specifically, which microscopic dynamics give rise to it? Is the hybridization with acoustic phonons relevant or just tangential to the BP problem? What sets its characteristic energy, and how does it evolve as the temperature increases toward the liquid state?

\section*{Discussion}
\subsubsection*{Transverse or longitudinal nature of the flat band}
A substantial body of evidence indicates that the boson peak is predominantly associated with transverse-like excitations.
First, across a wide range of glassy systems, the boson-peak frequency lies near the transverse Ioffe--Regel limit \cite{Shintani2008}, where transverse phonons begin to lose their plane-wave character.
This trend is also observed in all numerical models we have investigated (see SI), although to varying degrees.
Second, a more direct indication comes from the dynamical structure factor: the reduced inelastic intensity $B_T(q,\omega_{\rm BP})=\omega_{\rm BP}^2 S_T(q,\omega_{\rm BP})$ is typically much larger than its longitudinal counterpart $B_L(q,\omega_{\rm BP})$ (see SI), indicating that the excess modes responsible for the flat band couple more strongly to transverse than to longitudinal degrees of freedom.
Silica glasses, however, constitute a notable exception: a flat-band signal is clearly identifiable in both longitudinal and transverse channels.
As discussed in Ref.~\cite{mizuno2025bosonpeakcovalentnetwork}, this behavior has been attributed to silica’s proximity to the isostatic condition, which enhances the density of soft modes in both polarizations, and to strong non-affine effects that reduce the contrast between transverse and longitudinal responses.

Despite these preliminary interpretations, the polarization character of the flat band---whether fundamentally transverse, longitudinal, or mixed---remains insufficiently understood.
Clarifying this aspect is likely to be crucial for identifying its microscopic origin and for establishing a theoretical framework capable of capturing its degree of universality across different glass formers.

\subsubsection*{Universality of the BP in glasses and crystals} 
Beyond glassy systems, the boson peak anomaly has also been observed in a wide range of crystalline materials. 
Since the original proposal by Ackermann~\cite{PhysRevB.23.3886}, in crystals the most widely accepted theoretical explanation attributes this anomaly to the presence of exceptionally low-energy and abundant optical modes~\cite{PhysRevLett.93.245902,PhysRevB.70.212301,PhysRevB.99.024301,krivchikov2022role,doi:10.1142/S0217979221300024,Schliesser_2015,Baggioli_2020,mandyam2025glassy}. From a perspective that is agnostic of the microscopic details, as illustrated in Fig.~\ref{fig:0}, such low-energy optical modes bear striking resemblance to the flat BP band discussed in this work. 

Is this similarity just a coincidence? Could the flat BP band in glasses be understood as an ``optic-like'' collective excitation between soft and rigid regions? Is there a deeper universality linking the behavior of glasses and crystals? These are intriguing questions that warrant further exploration.

\section*{Conclusions}
In this work, we have compiled a broad range of evidence—both supporting and opposing—the scenario in which the boson-peak anomaly in amorphous solids is consistently associated with the appearance of a flat, weakly dispersive, non-phononic band in the dynamical structure factor. 
Alongside this survey of the literature, we presented new simulation results for two- and three-dimensional model glasses, as well as new analyses of experimental data from two-dimensional granular packings.

Taken together, these results provide strong support for a robust correspondence between the boson peak and the emergence of a flat band across a broad range of systems, including granular materials, polymers, silica, and metallic glasses. 
This correspondence already carries concrete implications for theory: as summarized in Table~\ref{nono}, several established approaches are in clear tension with the flat-band phenomenology and therefore require careful re-examination.
The central open questions now concern the microscopic origin of the flat band and the extent to which its characteristics—weak dispersion, intensity distribution, and polarization content—are universal across different classes of glasses.

\section*{Acknowledgments}
We thank Yuanchao Hu for collaboration at an early stage of this work. We are grateful to Giacomo Baldi for useful comments on a preliminary version of this manuscript.
We thank Huaping Zhang, Jack Douglas, Xun-Li Wang and Hajime Tanaka for inspiring discussions.
MB and CJ acknowledge the support of the Foreign Young Scholars Research Fund Project (Grant No.22Z033100604). MB acknowledges the sponsorship from the Yangyang Development Fund.
YJW was supported by the National Natural Science Foundation of China (Grant No. 12472112).
MPC acknowledges support by the Singapore Ministry of Education through grants MOE-T2EP50224-0016, RG152/23, and RG153/24. 
JZ acknowledges the support of the NSFC (Nos. 12534008, 11974238, and 12274291) and also the support of the Innovation Program of Shanghai Municipal Education Commission under No. 2021-01-07-00-02-E00138.

\onecolumngrid
\appendix 
\renewcommand\thefigure{S\arabic{figure}}    
\setcounter{figure}{0} 
\renewcommand{\theequation}{S\arabic{equation}}
\setcounter{equation}{0}
\renewcommand{\thesubsection}{SM\arabic{subsection}}

~\newpage

\section*{\Large Supporting Information}

\subsection*{Transverse dynamic structure factor}
\begin{figure}[!h]
    \centering
\includegraphics[width=0.6\linewidth]{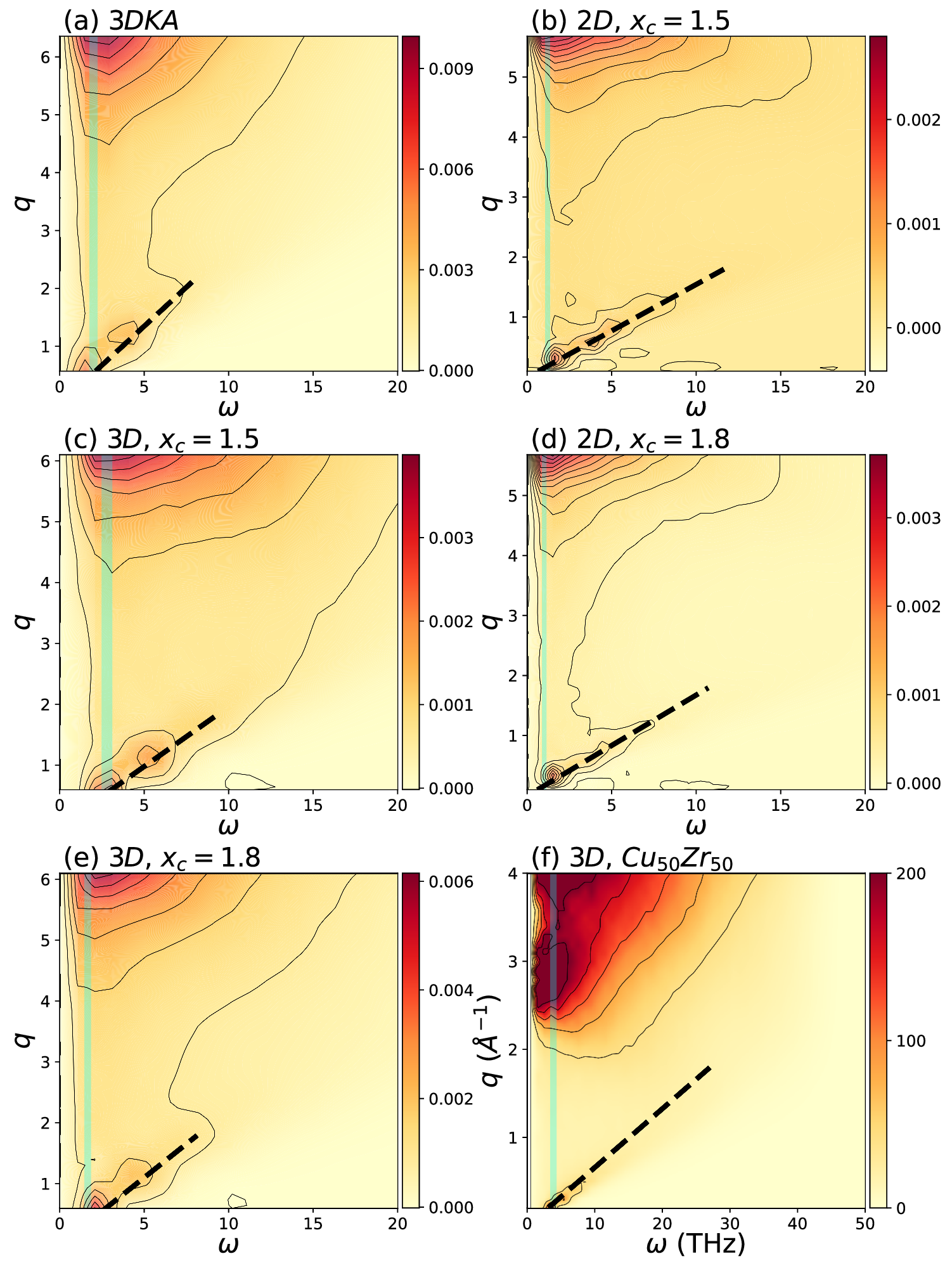}
    \caption{\color{blue}Two-dimensional contour plots for the transverse dynamic structure factor $S_T(q,\omega)$ for the 3D-KA binary mixture \textbf{(a)}; the 2D-attractive system with $x_c = 1.5$ \textbf{(b)}; the 3D-attractive system with $x_c = 1.5$ \textbf{(c)}; the 2D-attractive system with $x_c = 1.8$ \textbf{(d)}; the 3D-attractive system with $x_c=1.8$ \textbf{(e)}; the $\mathrm{Cu_{50}Zr_{50}}$ model \textbf{(f)}. Dashed lines correspond to the linear dispersion relation $\omega = vq$. The vertical bands identify the BP frequency extracted independently from the vibrational density of states.
    \label{fig:contour}
    }    

\end{figure}

Fig.~\ref{fig:contour} illustrates two-dimensional contour plots of the transverse dynamic structure factor $S_T(q,\omega)$ for all simulated systems.

The emergence of a flat, non-phononic band is clearly visible. Its frequency matches the independently determined boson peak frequency (indicated by the vertical cyan band), thereby confirming the findings reported in the main text.

~\newpage

\subsection*{Longitudinal dynamic structure factor}
\begin{figure*}[!h]
    \centering
    \includegraphics[width=\linewidth]{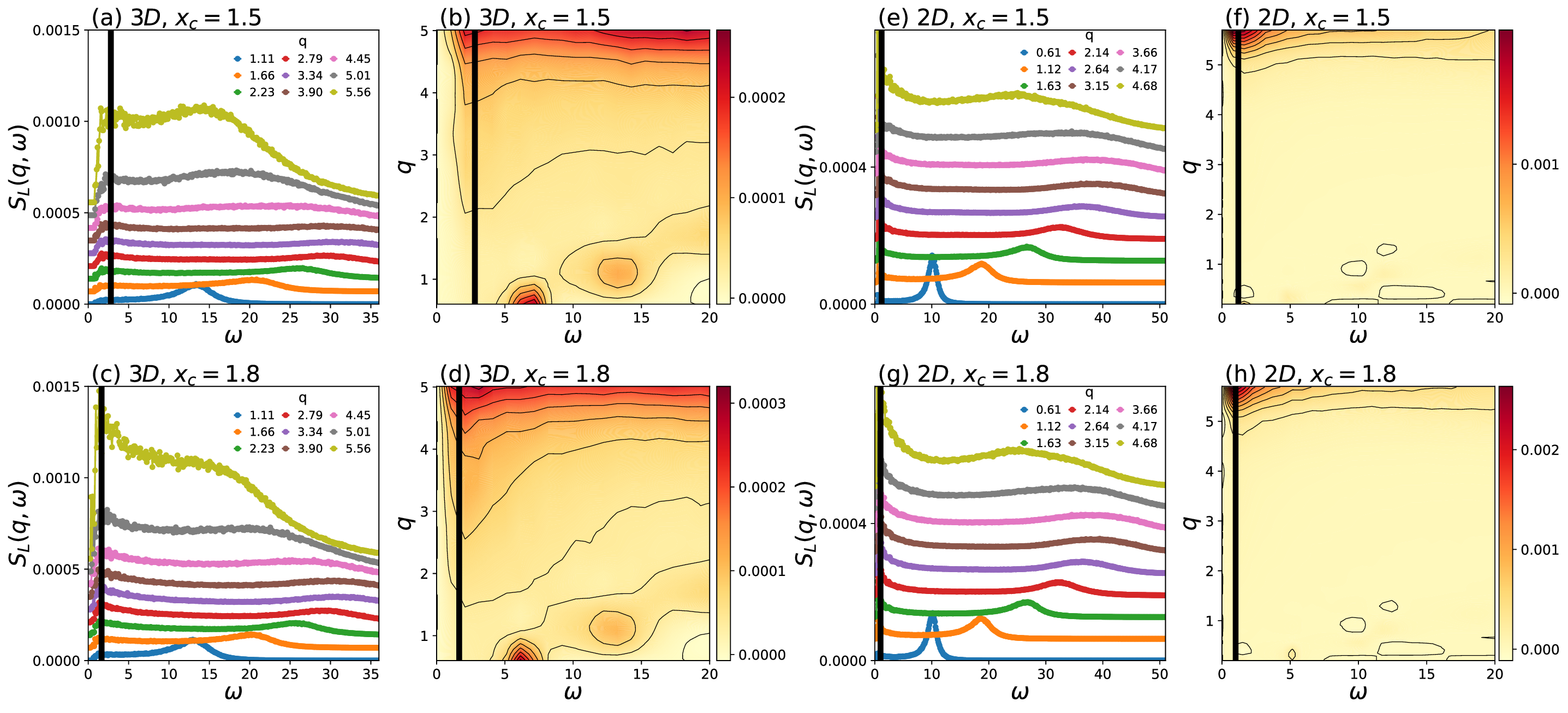}
    \caption{
    The longitudinal dynamic structure factor $S_L(q,\omega)$ for different wave vectors and its two-dimensional contour plot for the 3D-attractive system with $x_c = 1.5$ \textbf{(a)-(b)}; the 3D-attractive system with $x_c = 1.8$ \textbf{(c)-(d)}; the 2D-attractive system with $x_c = 1.5$ \textbf{(e)-(f)}; the 2D-attractive system with $x_c = 1.8$ \textbf{(g)-(h)}. The vertical bands identify the BP frequency extracted independently from the vibrational density of states.
    }
    \label{fig:review_numerics_longi}
\end{figure*}

In the main text, we presented evidence for the existence of a flat band in the transverse dynamic structure factor, which appears at the boson peak frequency. To explore whether this observation extends beyond the transverse sector, we here examine the longitudinal component, $S_L(q,\omega)$, as well.
Figure~\ref{fig:review_numerics_longi} displays $S_L(q,\omega)$ for some of the representative cases. 
Notably, we observe that a flat band emerges in the longitudinal spectrum as well, consistently with its observation in neutron-scattering experiments. 
The frequency of this flat band closely matches the boson peak frequency, mirroring the behavior observed in the transverse dynamic structure factor. 
This finding indicates that the flat-band feature is a robust aspect of the spectrum, influencing both transverse and longitudinal excitations. 
The presence of such a flat band in longitudinal modes reinforces the idea that the boson peak is associated with well-defined dispersionless excitation.

~\newpage
\subsection*{Linking Structure and Dynamics at the Boson Peak}
\begin{figure}[!h]
    \centering
    \includegraphics[width=0.7\linewidth]{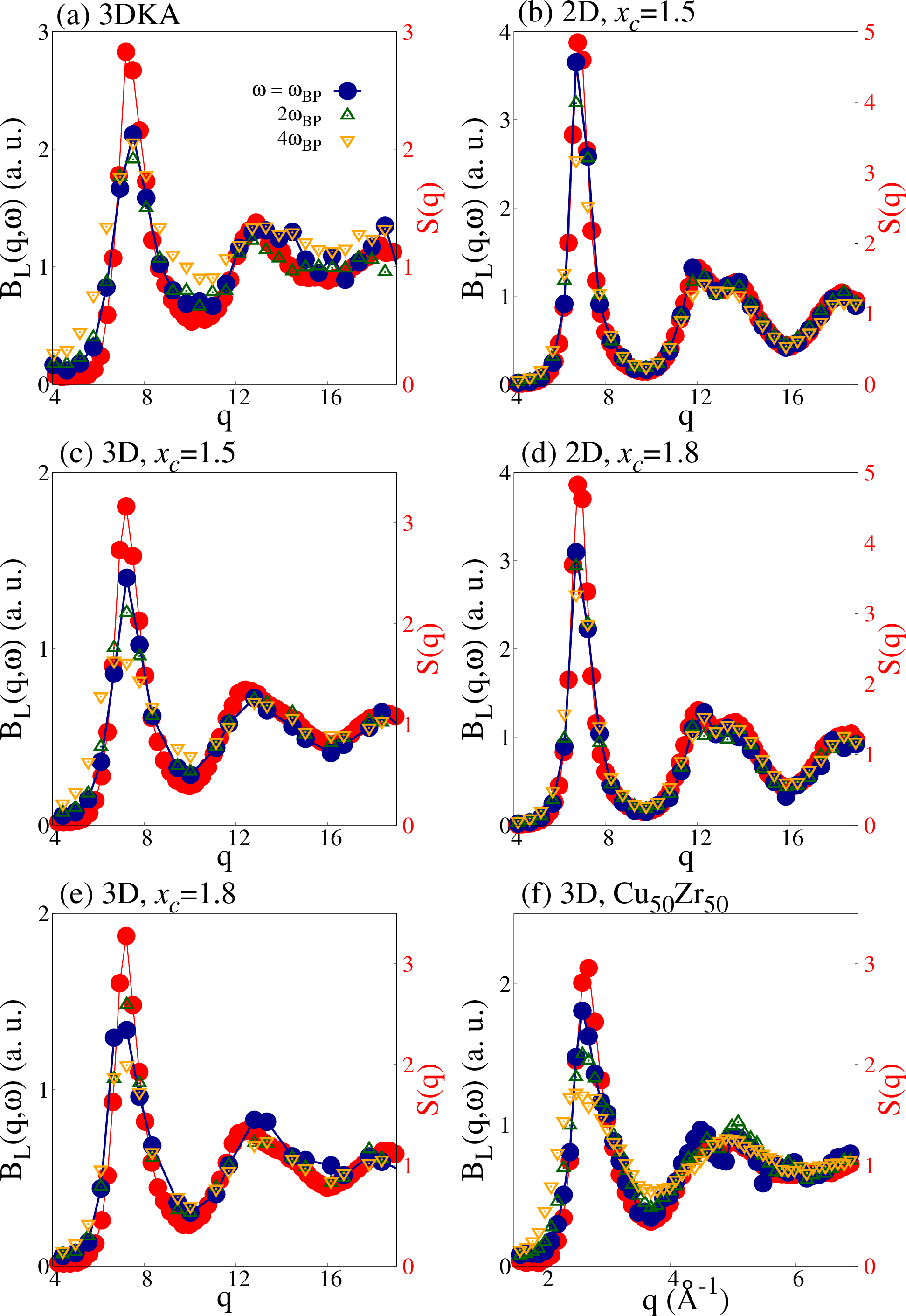}
    \caption{Comparison between the reduced dynamical structure factor states $B_L(q,\omega)$ calculated at different frequencies and the static structure factor $S(q)$ for different simulated systems.
    }    
    \label{corrL}
\end{figure}

In the main text, we analyzed the intensity of the flat-band signal in the transverse component using the reduced spectrum $B_T(q,\omega)$. This analysis revealed a strong correlation between $B_T(q,\omega)$ and the static structure factor, $S(q)$. Here, we extend this investigation to the longitudinal component. Figure~\ref{corrL} demonstrates that the reduced longitudinal spectrum, $B_L(q,\omega)$, exhibits a similarly striking, and in fact even stronger, correspondence with $S(q)$. These observations highlight a deep and consistent link between the static structure and dynamic response at the boson peak.

\subsection*{Transverse versus Longitudinal character of the Boson Peak}
\begin{figure}[!h]
    \centering
\includegraphics[width=0.5\linewidth]{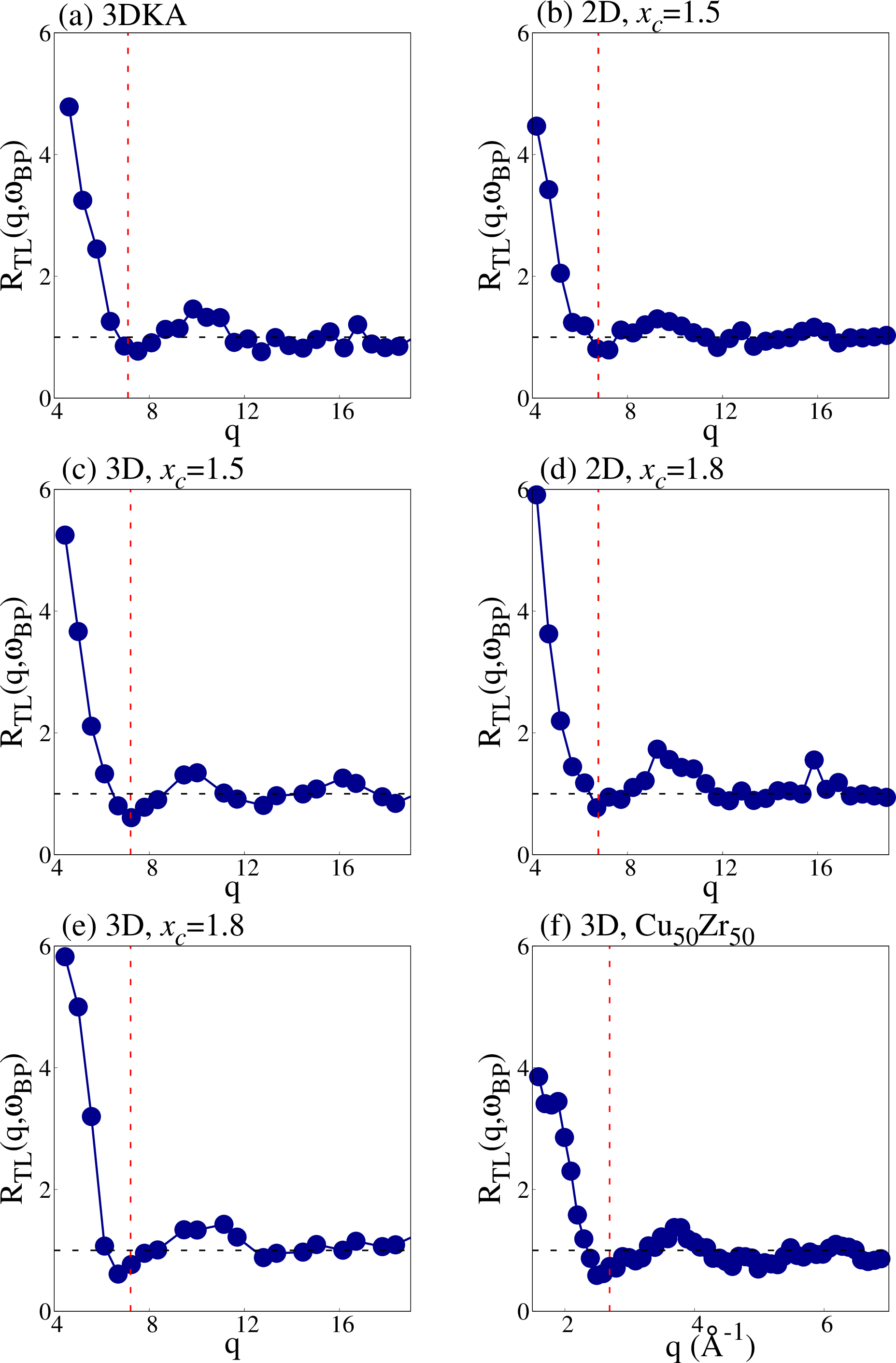}
    \caption{Wave-number dependence of the parameter $R_{TL}$ defined in Eq. \ref{eq:rtl}, which compares the relevance of longitudinal and transverse modes to the dynamical structure factor. 
    We evaluated $R_{TL}$ at the boson peak frequency for different simulated systems.
        \label{fig:ratio}
    }    

\end{figure}

To assess the transverse versus longitudinal character of the boson peak, we compare their
spectra at the boson-peak frequency. 
At sufficiently large $q$, modes at $\omega_{\mathrm{BP}}$ are strongly scattered and their polarization vectors can be approximated as isotropic. 
In $D$ spatial dimensions, an isotropic unit vector satisfies $\langle (\hat{\mathbf{q}}\!\cdot\!\mathbf{e})^{2} \rangle = 1/D$, yielding average longitudinal and transverse weights $1/D$ and $(D-1)/D$, respectively. 
One thus expects the limiting ratio $$\lim_{q \to \infty} S_T(q,\omega_{\mathrm{BP}})/S_L(q,\omega_{\mathrm{BP}}) \simeq D-1.$$
Henceforth, we measure the relative strength of the two components via the parameter 
\begin{equation}
R_{TL}(q,\omega_{\mathrm{BP}}) = \frac{1}{D-1}\,\frac{S_T(q,\omega_{\mathrm{BP}})}{S_L(q,\omega_{\mathrm{BP}})}
\label{eq:rtl}  
\end{equation}
Figure~\ref{fig:ratio} shows that for $q \geq q^\ast$, $R_{TL}(q,\omega_{\mathrm{BP}}) \simeq 1$, indicating that large-$q$ polarizations at the BP frequency are essentially isotropic. 
In contrast, $R_{TL}(q,\omega_{\mathrm{BP}})$ increases for $q < q^\ast$, demonstrating that the more extended modes contributing to the BP retain a predominantly transverse character on the structural length scale.

~\newpage
\subsection*{The connection with the Ioffe-Regel limit}
Motivated by previous studies suggesting a close relation between the boson peak and the transverse Ioffe–Regel crossover, we examine whether the boson-peak frequency correlates with the Ioffe–Regel limit $\omega_{\mathrm{IR}}$ for transverse acoustic phonons.  
The latter is defined as the frequency at which propagating {\it transverse} phonons become overdamped, i.e. $\Omega(q)=\pi\Gamma(q)$, with $\Omega$ and $\Gamma$ the energy and linewidth of the transverse acoustic mode.

We summarize our results in Table~\ref{tab:bp_ir}. Taken together, these comparisons suggest that the boson-peak and transverse Ioffe–Regel scales track each other and are generally close, particularly in three dimensions. Whether this correspondence between the BP and the IR scales is related somehow to the existence of the flat band discussed in our \textit{Perspective} remains an open question.

\begin{table}[!h]
\centering
\begin{tabular}{lcccc}
\hline
System & $\omega_{\mathrm{BP}}$ & $\omega_{\mathrm{IR}}$ &
$\omega_{\mathrm{BP}}/\omega_{\mathrm{IR}}$ & Type \\
\hline
2D, $x_c = 1.5$                & 1.2  & 2.9  & 0.41 & Numerical \\
2D, $x_c = 1.8$                & 1.0  & 1.9  & 0.52 & Numerical \\
2D, Photoelastic disks(Fig.5) & 0.43 & 0.41 & 1.04 & Experimental \\ 
3DKA                           & 2.0  & 2.1  & 0.95 & Numerical \\
3D, $x_c = 1.5$                & 2.8  & 3.4  & 0.82 & Numerical \\
3D, $x_c = 1.8$                & 1.65 & 2.3  & 0.72 & Numerical \\
3D, Cu$_{50}$Zr$_{50}$         & 3.8 & 3.2  & 1.18 & Numerical \\
3D, IPL (Fig.2) & 2.46 & 2.55 & 0.96 & Numerical \\
3D, Silica (Fig.3) & 1.5 & 1.45 & 1.03 & Experimental \\

\hline
\end{tabular}
\caption{Boson-peak frequency $\omega_{\mathrm{BP}}$, Ioffe–Regel frequency 
$\omega_{\mathrm{IR}}$, their ratio, and whether the data are numerical or experimental.}
\label{tab:bp_ir}
\end{table}

\end{document}